\documentclass{aa}  

\usepackage{graphicx}
\usepackage{txfonts}

\usepackage{float}
\usepackage{caption}
\usepackage{subcaption}
\usepackage{booktabs}
\usepackage{multirow}
\usepackage{multicol}
\usepackage{makecell}
\newcommand{\ra}[1]{\renewcommand{\arraystretch}{#1}}
\usepackage{lscape}
\usepackage{caption}
\usepackage{subcaption}
\usepackage{graphicx}
\def\nbR{\ensuremath{\mathrm{I\! R}}}
\def\nbE{\ensuremath{\mathrm{I\! E}}}

\usepackage{xcolor}
\usepackage{ulem}

\begin{document} 

   \title{Unsupervised classification of CIGALE galaxy spectra}
   
   
   \author{J. Dubois\inst{1}
        \and
        D. Fraix-Burnet\inst{1}
        \and
        J. Moultaka\inst{2}
        \and
        P. Sharma \inst{3}
        \and
        D. Burgarella \inst{3}
   }
   
   \institute{Univ. Grenoble Alpes, CNRS, IPAG, Grenoble, France \\ \email{julien.dubois@univ-grenoble-alpes.fr, didier.fraix-burnet@univ-grenoble-alpes.fr}
        \and   IRAP, Université de Toulouse, CNRS, CNES, UPS, 14, avenue Edouard Belin, F-31400 Toulouse, France\\ \email{jihane.moultaka@irap.omp.eu}
        \and Aix-Marseille Université, CNRS, LAM (Laboratoire d’Astrophysique de Marseille) UMR 7326, 13388 Marseille, France  \\ 
   }
   
   \date{Received March 15, 2021; accepted}
   
\abstract
{}
{Our study aims at providing deeper insight into the power and limitation of an unsupervised classification algorithm (called Fisher-EM) on spectra of galaxies. This algorithm uses a Gaussian mixture in a discriminative latent subspace. To this end, we investigate the capacity of this algorithm to segregate the physical parameters used to generate mock spectra and the influence of the noise on the classification.
}
{With the code CIGALE and different values for nine input parameters characterising the stellar population, we simulated a sample of 11 475 optical spectra of galaxies containing 496 monochromatic fluxes. The statistical model and the optimum number of clusters are given in Fisher-EM by the integrated completed likelihood (ICL) criterion. We repeated the analyses several times to assess the robustness of the results. }
{Two distinct classifications can be distinguished in the case of the noiseless spectra. The classification with more than 13 clusters disappears when noise is added, while the classification with 12 clusters is very robust against noise down to a signal-to-noise ratio (S/N) of 3. At S/N=1, the optimum is 5 clusters, but the classification is still compatible with the previous classification. The distribution of the parameters used for the simulation shows an excellent discrimination between classes. A higher dispersion both in the spectra within each class and in the parameter distribution leads us to conclude that despite a much higher ICL, the classification with more than 13 clusters in the noiseless case is not physically relevant.}
{This study yields two conclusions that are valid at least for the Fisher-EM algorithm. Firstly, the unsupervised classification of spectra of galaxies is both reliable and robust to noise. Secondly, such analyses are able to extract the useful physical information contained in the spectra and to build highly meaningful classifications. In an epoch of data-driven astrophysics, it is important to trust unsupervised machine-learning approaches that do not require training samples that are unavoidably biased.}

   \maketitle
   
   \section{Introduction}
   \label{section:introduction}
   
  Machine learning is becoming increasingly popular in astrophysics mainly through the supervised approach, which consists of training the algorithm with the relevant information we already know. This is for instance the case of the classification of astronomical objects, where the observations are matched against representative data of previously established classes \citep{Fraix-Burnet2015}. 
   
   Supervised classification has two appealing advantages. The first advantage is that it can be very fast and is well adapted to the very large databases produced by new telescopes. The second advantage is its immediate usefulness since a new observation is assigned to a class with supposedly known physical properties.
   
   Supervised classification has several limitations, however. It depends much on the quality of the reference classification, and this relies on important prerequisites: data samples that are large enough, good quality of the data, and a proper understanding of the physics underlying the studied objects. Moreover, a careful visual inspection may be needed in many cases. Hence, rarer objects might not be well represented, inducing biases in the learning process \citep[e.g.][]{Cavuoti2014}. Supervised classification is also obviously not suited to characterising or identifying new types of objects. 
   
   Conversely, unsupervised classification consists of pattern recognition in the data space to establish a reference classification. Cleared from human subjectivity, this approach can be expected to be more suited for subsequent supervised classification since it is entirely data driven.
   
  Among clustering techniques, model-based approaches \citep{Fraley2002,McLachlan2000} are popular. They are renowned for their probabilistic foundations and their flexibility. One of the main advantages of these approaches is the fact that their models and results can be interpreted from both the statistical
and practical points of view. In addition, many of the other heuristic approaches (mostly based on similarity measures) approximately correspond to particular clustering models \citep{Bouveyron2019}. One of the simplest and well-known algorithms is the k-means approach, which considers a mixture of identical Gaussians. 
A more powerful tool is the Gaussian mixture model \citep[GMM; ][]{Bouveyron2019} approach, which fits a multivariate Gaussian to each cluster \cite[e.g.][]{Souza2017}.
The GMM allows for more adaptability to the distribution of points and clusters in the data space, but that may not be sufficient. Two solutions are possible: merging of Gaussian components \citep[e.g.][]{Hennig2010}, and use of non-Gaussian components \citep[see a comprehensive review in][]{Bouveyron2019}. However, the choice of the merging criterion and of the models raises philosophical issues \citep[e.g.][]{Hennig2015}. In our experience, if the data do not fit the model, the algorithm fails to find the optimum number of clusters, as in \citet{De2016}. We never met this problem with GMM on astrophysical data. 

Unfortunately, model-based methods usually show a disappointing behaviour in high-dimensional spaces \citep[e.g.][]{De2016}. They suffer from the well-known curse of dimensionality \citep{Bellman2010}, which is mainly due to the fact that model-based techniques are over-parametrised in high-dimensional spaces. For this reason, dimension reduction methods are frequently used in practice to reduce the dimension of the data before the clustering step. Feature extraction methods, such as principal component analysis (PCA), or feature selection methods are very popular. In astrophysics, PCA is used to separate some large classes of spectra or even remove some noise \citep{Marchetti2013}. However, dimension reduction usually does not consider the clustering task and provides a suboptimal data representation for the classification step. For instance, the variance axes of PCA are not necessarily the discriminant axes \citep{Chang1983}. Dimension reduction methods usually imply an information loss that could have been discriminative. To avoid the drawbacks of dimension reduction, several approaches have been proposed in the past decade to allow model-based methods to efficiently classify high-dimensional data. Subspace clustering techniques are one such approach. These techniques are mostly based on probabilistic versions of the factor analysis model and allow classifying the data in low-dimensional subspaces without reducing the dimension \citep{Bouveyron2016}.

The main drawback of unsupervised classification is that the classes are built from a statistical point of view. Statistical criteria (and this is the advantage of model-based methods) provide objective ways to find the 'best' solution and assess their robustness. But there is no real 'good' classification in the sense that only the physical interpretation and the goal of the study constitute the metric assessing the usefulness of the classification \citep[e.g.][]{Hennig2015}. Nevertheless, several works have proven the relevance of this approach in astrophysics  \citep[see a review for the extragalactic domain in][]{Fraix-Burnet2015}. For instance, \citet{Souza2017} used a GMM \citep[e.g.][]{Bouveyron2019} approach to classify emission line galaxies, using as observables the two line ratios $\log$[OIII]/H$\beta$ and $\log$[NII]/H$\alpha$ and the equivalent width of the H$\alpha$ line $(\log$EW(H$\alpha$)). They reported a statistical optimum of four classes that slightly revised the separation between active galactic nuclei (AGNs), Seyfert galaxies, low-ionisation nuclear emission-line regions (LINERs), and star-forming regions. With a larger sample of 362923 galaxies and 47 observables (including emission lines, Lick indices, morphologies, and photometric observables), \citet{Chattopadhyay2019} applied an independent component analysis \citep{Jutten1991} to reduce the dimensionality, followed by a k-means \citep{kmeans1967} analysis. They obtained ten classes that correspond to the classically known classes of galaxies.
   
   A significant step forward in both the sample size and the number of features has been performed by \citet{Fraix-Burnet2021} on a sample of 702248 optical spectra (1437 monochromatic fluxes each) of galaxies from the Sloan Digital Sky Survey (SDSS) using an unsupervised clustering discriminative latent mixture model algorithm called Fisher-EM \citep{Bouveyron2012Jan}. They obtained 86 robust and very homogeneous classes for which a preliminary analysis shows that they can be easily given a physical interpretation.

   In the present paper, we aim at bringing some insight into the power and limitation of the use of the algorithm Fisher-EM on a simple mock sample of spectra. This is a follow-up of the above study \citep{Fraix-Burnet2021}, but its results can be helpful for many unsupervised approaches. It is well known that the many degeneracies present in spectra cause the derivation of the properties of galaxies to be somewhat difficult \citep{Marchetti2013}. We are particularly interested in the way the classification obtained through Fisher-EM is able or fails to segregate the physical properties and in its sensitivity to noise.  
 
   To this end, we simulated a sample of galaxy spectra with the spectral energy distribution (SED) fitting code CIGALE \citep{Boquien2019} using a number of input physical parameters. Then we performed the clustering with Fisher-EM on the spectra. Finally, we analysed the resulting classes in terms of these input parameters. We also study the influence of the noise on the results.
   
   This paper is organised as follows. The simulations and the data are presented in Sect.~\ref{section:data}. The algorithm Fisher-EM is briefly described in Sect.~\ref{section:method} together with the method for selecting the optimum number of clusters. In Sect.~\ref{section:noiseless} we describe the classifications obtained on the noiseless spectra, and we show the distribution of the physical parameters of the simulations and determine the most discriminant distribution. In Sect.~\ref{section:noise}, the influence of different levels of noise on the classification and on the associated physical properties of the classes is presented. After a discussion in Sect.~\ref{section:discussion}, we conclude this paper in Sect.~\ref{section:conclusion}.

\section{Data}

\label{section:data}

\subsection{Generation of the spectra}

    \begin{table}
        \centering
        \caption{Parameter linear correlation coefficients. The parameters are not intrinsically correlated in CIGALE, but the combinations of values used to generate the sample for this study may show underlying involuntary correlations between some parameters.}
        \resizebox{\columnwidth}{!}{
        \ra{1.3}
        \begin{tabular}{r r r r r r l}
                \hline\hline    
             & T$_{main}$ & $\tau_{main}$ & f$_{burst}$ & T$_{burst}$ & $\tau_{burst}$ & Metallicity\\
             \hline
                \cline{1-7}
             $\tau_{main}$ & $0.17 $&  & & & &  \\
             f$_{burst}$ & $-0.42$ & $0.01$ &  & & &  \\
             T$_{burst}$ & $0.21$ & $0.18$ & $-0.28$ &  & &  \\
             $\tau_{burst}$ & $-0.06$ & $0.38$ & $-0.11$ & $0.20$ &  &  \\
             Metal. & $0.33$ & $-0.20$ & $-0.39$ & $0.11$& $-0.02$ &   \\
             E(B-V)$_{cont}$ & $-0.22$ & $0.10$ & $0.27$ & $-0.10$ & $-0.01$ & $-0.28$   \\
             \hline
        \end{tabular}}
        
        \label{table:correlation}
        \end{table}

 The spectra used in this study were simulated with CIGALE (Code Investigating GALaxy Emission), a software that extends the work of \cite{Burgarella2005Jul} and \cite{Noll2009Dec}. It creates spectra from the UV to the far-IR. We consider 
   the optical part of the spectra (496 fluxes between 380.66-737.00~nm)
   to be able to make a later comparison with observed data such as those from the SDSS \citep[see e.g.][]{Fraix-Burnet2021}. Each simulated spectrum requires a set of input parameters characterising the galaxy.  
     In our simulations, we assumed a \citet{Chabrier2003} initial mass function (IMF). We limited the star formation history (SFH) to a main population and in some cases, to a later burst of the same metallicity. With these assumptions, the optical part of the spectra is affected by nine parameters in CIGALE. We simulated spectra with bursts of varying strength (f$_{burst}$): none, medium, and large. For each burst strength, we considered populations of different ages and timescales (T$_{main}$, $\tau_{main}$, T$_{burst}$, $\tau_{burst}$), metallicity, reddening (E$_{(B-V)lines}$, E$_{(B-V)factor}$), and redshift. See Sect.~\ref{section:data:params} for the description of the input parameters.
      The combinations of the values chosen 
      for the input parameters yield 11475 spectra. All spectra were normalised by their mean values between 505 and 581~nm, a region where the spectra have no emission lines.

   \subsection{Description of the input parameters}
   \label{section:data:params}
\begin{figure}
        \centering
        \includegraphics[width=\hsize]{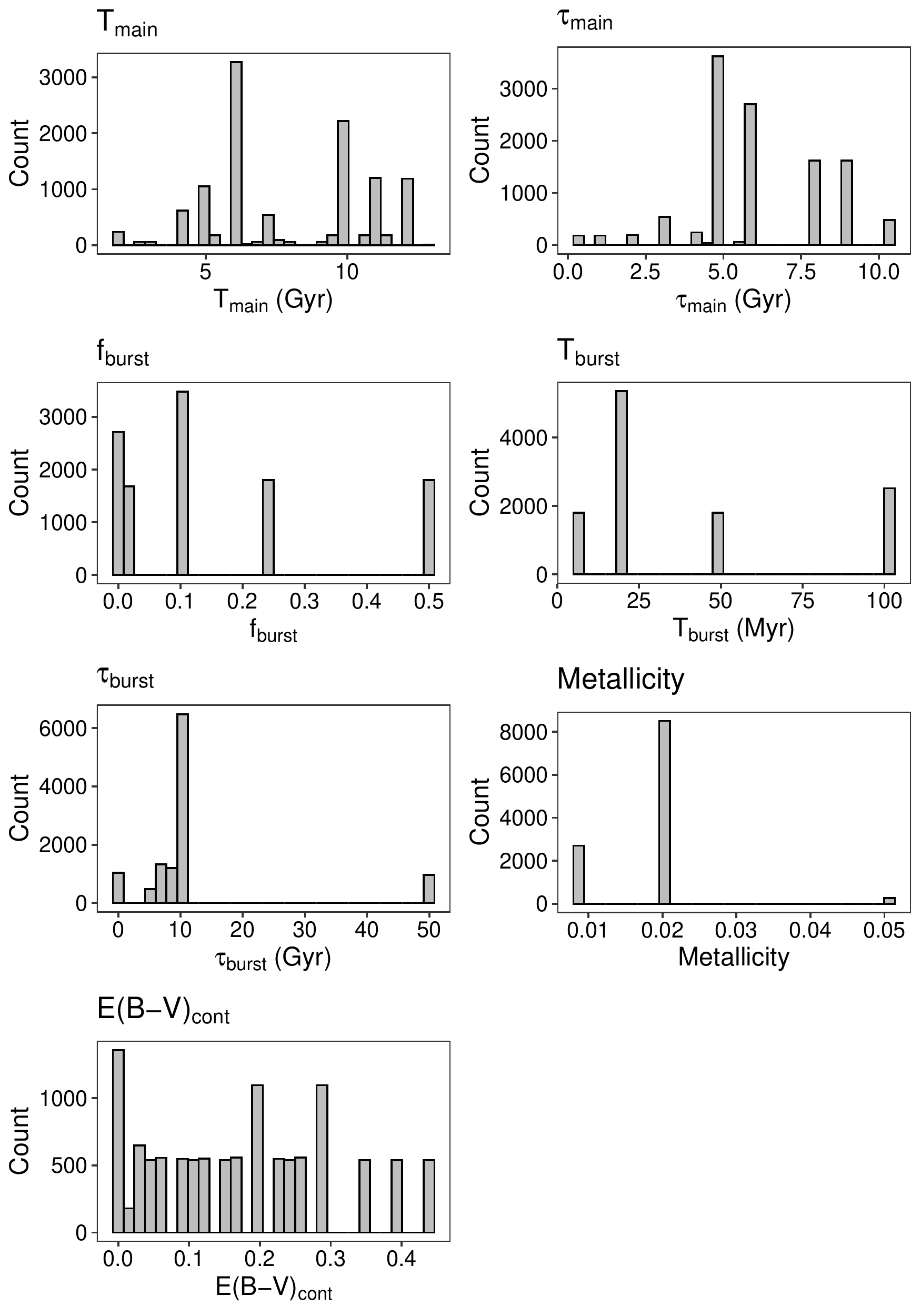}
        \caption{Histograms of CIGALE parameters input values in the study sample. This sample was not created to fit some specific parameter distribution, but rather covers the parameter space as much as CIGALE allowed it while keeping the spectra realistic.}
        \label{fig:paramsdistribs}
\end{figure}

    The optical part of the spectra is affected by nine CIGALE input parameters. The values taken in this work are listed in Table~\ref{table:parameters} and their distribution is shown in Fig.~\ref{fig:paramsdistribs}. These nine parameters are reduced to seven, as explained below.
   
    T$_{main}$ is the age (in Myr) of the main stellar population in the galaxy. It is varied from 2000 to 13000 Myr. $\tau_{main}$ is the e-folding time (in Myr) of the main stellar population and characterises the star formation rate (SFR) of the main population together with T$_{main}$ (module \textit{sfhdelayed} of CIGALE), 
    \begin{equation}
        \textrm{SFR}_{main}(t) \propto \frac{t}{\tau_{main}^2} \exp(-t/\tau_{main}) 
    .\end{equation}
    In the sample, $\tau_{main}$  varies from 500 to 10500 Myr. f$_{burst}$ is the mass fraction of stars produced during a burst of star formation, and ranges from 0 to 0.5. When f$_{burst}=0$, no burst is considered in the history of the galaxy. T$_{burst}$ is the age (in Myr) of the burst of star formation, and ranges from 5 to 100 Myr. When f$_{burst}=0$, T$_{burst}$ is fixed to 0. $\tau_{burst}$ is the e-folding time (in Myr) of the burst of star formation and characterises the SFR of the burst event together with T$_{burst}$,
    \begin{equation}
        \textrm{SFR}_{burst}(t) \propto \frac{t}{\tau_{burst}^2} \exp\left(\frac{-t}{\tau_{burst}}\right) \text{\quad for } t>T_{main} - T_{burst} 
    .\end{equation}
    It varies from 4500 to 50000 Myr. These values are of the same order as T$_{main}-$T$_{burst}$ and much higher than T$_{burst}$, so that the SFR is essentially constant during the second burst of star formation.  When f$_{burst}=0$, no burst is considered in the history of the galaxy, and $\tau_{burst}$ is therefore fixed to 0.  Metallicity is assumed to be identical for the main and burst stellar populations as well as for the interstellar medium, and it takes three possible values: 0.008, 0.02, and 0.05. E(B-V)$_{cont}$ is the reddening of the continuum that intervenes in the computation of the attenuation due to the dust inside the galaxy. For the lines, we have,
    \begin{equation}
        E(B-V)_{lines}=E(B-V)_{cont}/E(B-V)_{factor}
    \end{equation}
    where E(B-V)$_{factor}$ is a normalising factor given as an input parameter. We choose to discuss only E(B-V)$_{cont}$ throughout this paper for its physical relevance. In the simulated spectra, E(B-V)$_{cont}$ takes a wide variety of values from 0.000125 to 0.44.
    Redshift was varied from 0 to 1 to include the effect of the intergalactic medium (IGM) on the spectra. For the classification process, we shifted the spectra to the same redshift of~0. As a consequence, the spectra that differ only by redshift end up being very similar because the effect of the IGM happens to be very small. We therefore do not discuss the redshift in the remainder of this paper.

As defined previously, the T$_{main}$, $\tau_{main}$, f$_{burst}$, T$_{burst}$ , and metallicity parameters are the building blocks of the stellar populations composing a galaxy. It is thus natural that varying the values of these parameters directly influences the continuum and the absorption lines of a galaxy spectrum (e.g. a high value of T$_{main}$ reddens the spectrum and deepens molecular lines). Moreover, E(B-V)$_{cont}$ translates the presence of dust into the 
interstellar medium by reddening the spectrum. On the other hand, the emission lines in a galaxy spectrum are the signature of a recent star-forming event in the history of that galaxy. Therefore, the same parameters T$_{main}$, $\tau_{main}$, f$_{burst}$, and T$_{burst}$ are also responsible for the presence and the height of these lines. Moreover, since the interstellar medium in CIGALE has the same metallicity as the stars, the metallicity parameter will also affect the emission lines in addition to the stellar component. Finally, because the chosen values of $\tau_{burst}$ are much higher than those of T$_{burst}$
(see Table~\ref{table:parameters}), $\tau_{burst}$ only weakly affects our spectra.

Most CIGALE parameters only take discrete and pre-determined values, making it impossible to fully cover the parameter space. Some parameters such as E(B-V)$_{cont}$ allow for a decent sampling, while some others such as metallicity only take very few values. The sampling density of a parameter may affect its discriminant power slightly, but in a very indirect way since the clustering is performed with the spectra and not with the input parameters.
    
Moreover, while the parameters are not intrinsically correlated in CIGALE, some of them may be slightly correlated due to our parameter sampling. Such correlations have to be kept in mind when analysing the discriminative properties of the classification method, as one parameter may deceptively appear well discriminated due to its correlation with another discriminated parameter. However, for our sample, the Pearson correlation coefficients remain small (see Table~\ref{table:correlation}) and reaches -0.42 (f$_{burst}$ versus T$_{main}$) at most.
   
  We wish to insist that our mock sample is not intended to represent a complete diversity of spectra of real galaxies. Rather, it must be considered as a realistic sample with selection biases. These biases are here not observational, instrumental, or catalogue based, but are due to the necessarily limited choice for the values of the input parameters. 
   
   The simulated spectra are noiseless, and their analysis is presented in Sect.~\ref{section:noiseless}. To study the influence of the level of noise on the clustering result, we generated sets with different signal-to-noise ratio (S/N) spectra by adding a Gaussian noise to each monochromatic flux. The analysis of these noisy spectra is described in Sect.~\ref{section:noise}. 
   
\section{Method}
    \label{section:method}
   \subsection{Fisher-EM algorithm}
   We have applied the unsupervised classification method called Fisher-EM \citep{Bouveyron2012Jan} on the sample of optical galaxy spectra simulated with CIGALE (see Sect. \ref{section:data}). Fisher-EM is a subspace Gaussian mixture algorithm that relies on a statistical model, called the discriminative latent mixture (DLM) model. It uses a modified version of the expectation-maximisation (EM) algorithm by inserting a Fisher step to optimise the ratio of the sum of the between-class variance over the sum of the within-class variance for a better clustering.
   
   Formally, we may define the observation vector $\bf{Y}=\{y_1, ..., y_n\}$ such that $y_i\in \nbR^{p}$ describes spectrum number $i$. The dimension $p$ is the $p$ fluxes at the $p$ wavelengths of the spectra.
  
  Classifying the observations into K classes mathematically translates into finding the vector $\bf{Z}=\{z_1, ..., z_n\}$ , which assigns each spectrum $y_i$ to a given class $z_i \in  [\![1,K]\!]$. 
  In the case of Fisher-EM, the clustering process occurs in a subspace $\nbE \subset \nbR^p$ of dimension $d=K-1 < p$. 
Therefore, the Gaussian mixture model is applied to the projected data $\bf{X}$ rather than the observed data $\bf{Y}$,
   \begin{equation}
   \centering
       \bf{Y} = \bf{U}\bf{X} + \bf{\epsilon}
       \label{eq:Y}
   ,\end{equation}
   where $\bf{U} \in \mathcal{M}_{p,d}(\nbR)$ is the projection matrix and $\bf{\epsilon}$ is a noise vector of dimension $p$ following a Gaussian distribution centred around 0 and of covariance matrix $\Psi$ ($\varepsilon_{k}\sim\mathcal{N}(0,\Psi_{k})$). The multivariate Gaussian probability distribution $\bf{X}$ describing the class $k$ in the subspace is parametrised by a mean vector $\bf{\mu_k}$ and a covariance matrix $\bf{\Sigma}_k$,
   \begin{equation}
       \centering
       X|_{Z=k} \sim \mathcal{N}(\mu_k, \Sigma_k)
       \label{eq:X}
   .\end{equation}
   Combining Eqs.~\ref{eq:Y} and \ref{eq:X}, we obtain
   \begin{equation}
       \centering
       Y_{|X,Z=k}\sim\mathcal{N}(UX,\Psi_{k})
   .\end{equation}
   The observed data are thus modelled by a marginal distribution $f(\bf{y})$ that is the sum of K multivariate Gaussian density functions $\phi$ of mean $\bf{U}\mu_k$ and covariance $\bf{U}\Sigma_k\bf{U^t} + \Psi$, each weighted by the corresponding mixing proportion $\pi_k$,
    \begin{equation}
       \centering
       f(\bf{y}) = \sum_{k=1}^K \pi_k \phi(y;\bf{U}\mu_k, \bf{U}\Sigma_k\bf{U^t} + \Psi)
\label{eq:f}
   .\end{equation}
   
   By further assuming that the noise covariance matrix $\Psi_{k}$
satisfies the conditions $V^{t}\Psi_{k} V=\beta_{k}\mathbf{I}_{p-d}$ , where $V$ is the orthogonal complement of $U$, and $U^{t}\Psi_{k} U=\mathbf{0}_{d}$, the whole statistical model denoted by $\mathrm{DLM}_{[\Sigma_{k}\beta_{k}]}$
can be shown to take the following form: $$ \left(  \begin{array}{c@{}c} \begin{array}{|ccc|}\hline ~~ & ~~ & ~~ \\  & \Sigma_k &  \\  & & \\ \hline \end{array} & \mathbf{0}\\ \mathbf{0} &  \begin{array}{|cccc|}\hline \beta_{k} & & & 0\\ & \ddots & &\\  & & \ddots &\\ 0 & & & \beta_{k}\\ \hline \end{array} \end{array}\right)  \begin{array}{cc} \left.\begin{array}{c} \\\\\\\end{array} \right\}  & d \leq K-1\vspace{1.5ex}\\ \left.\begin{array}{c} \\\\\\\\\end{array}\right\}  & (p-d)\end{array}$$ These
last conditions imply that the discriminative and the non-discriminative
subspaces are orthogonal, which suggests in practice that all the
relevant clustering information remains in the latent subspace. From a practical point of view, $\beta_{k}$ models the variance of the non-discriminative noise of the data.
   
   Several other models can be obtained from the DLM$_{[\Sigma_{k}\beta_{k}]}$
model by relaxing or adding constraints on model parameters. It can, for example, be assumed that the noise parameter $\beta_{k}$ differs from class to class, or that the covariance matrices $\Sigma_k$ are the same for all $K$ classes.
   A thorough description of the DLM model, its 12 declinations, and the algorithm itself can be found in \cite{Bouveyron2012Jan}.
   
   The Fisher-EM algorithm requires the number of groups K and the DLM model as input.
   After undergoing an initialisation obtained from multiple k-means runs, the algorithm proceeds as follows:
   \begin{itemize}
       \item E-step: The posterior probabilities that the $n$ observations $y_i$ belong to each of the K classes are computed.
       \item Fisher-step: 
       The projection matrix U is computed to maximise the Fisher criterion.
       \item M-step: The DLM model parameters are adjusted to maximise the likelihood.
   \end{itemize}
   
   The Fisher-EM algorithm is implemented in the eponym package for \textsf{R}.
   
   \subsection{Choice of the model and number of classes}
   
   The choice of the best statistical DLM model and the optimum number of clusters depends on the data and was estimated with the integrated completed likelihood (ICL) criterion. This criterion  penalises the likelihood by the number of parameters of the statistical model, the number of observations, and favours well-separated clusters \citep{Biernacki2000,Girard2016}. 
   
  The best statistical model was found to be $A_{kj}B_k$ in all the cases studied in this paper. This model is such that in each group, the covariance matrix $\bf{\Sigma}_k$ is assumed to be diagonal: $\bf{\Sigma}_k = diag(\alpha_{k_1},...,\alpha_{k_d})$.
The optimum number of clusters K is also given by the maximum ICL value and depends on each data set.
   
   In order to characterise the stability of the algorithm, several classifications were generated for each number of clusters K, yielding a distribution of ICL values for each. This dispersion is found to be generally low and is shown as boxplots in the ICL-versus-K figures presented throughout this paper.
   
   The EM algorithm is known to sometimes find an empty cluster (i.e. a null $\pi_k$ in Eq.~\ref{eq:f}). This results in an undefined log-likelihood that stops Fisher-EM. In this paper, we call this the non-convergence of the algorithm.

\section{Analysis of the noiseless spectra}
\label{section:noiseless}

\subsection{Optimal number of clusters}
\label{section:noiseless:nbclusters}
\begin{figure}
        \centering
        \includegraphics[width=\hsize]{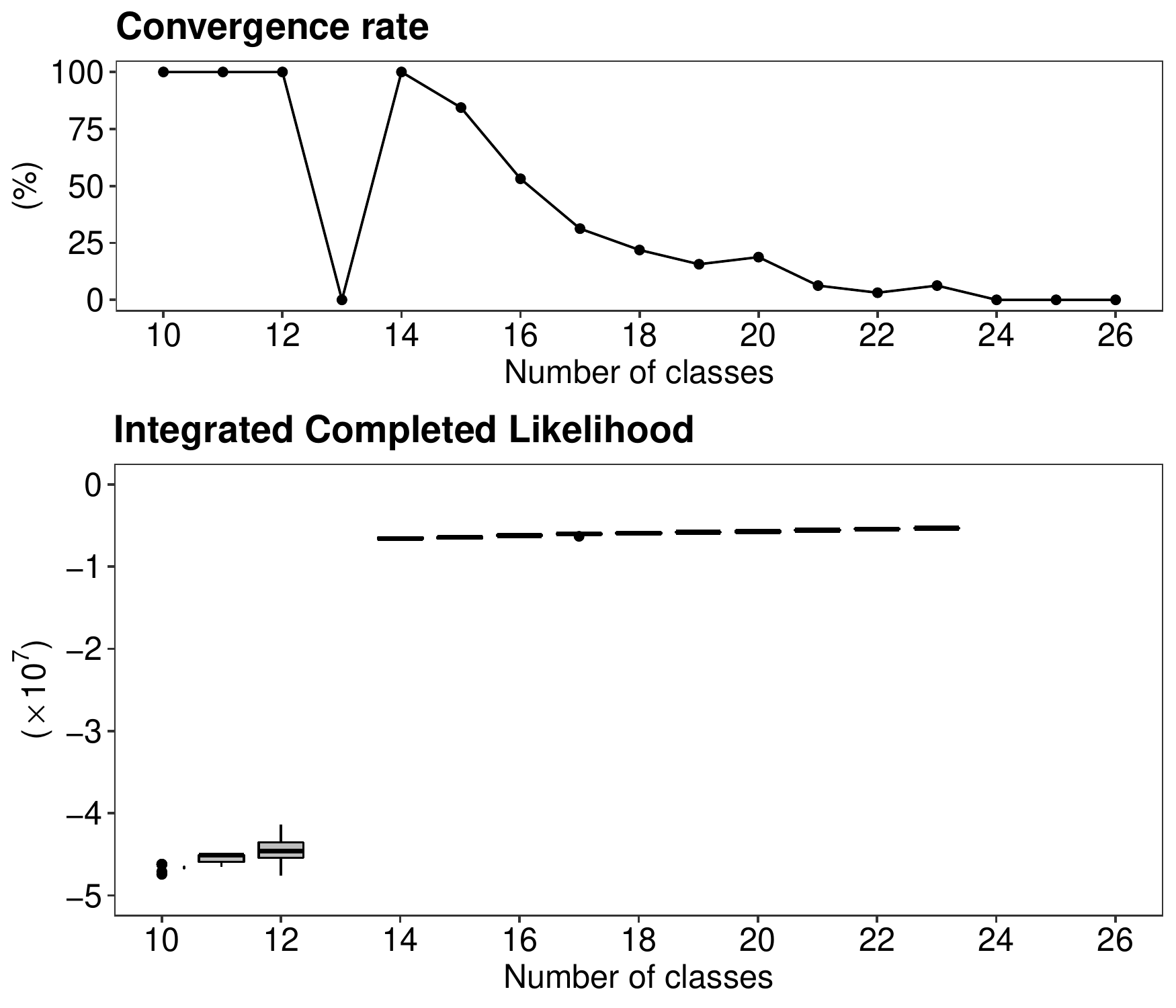}
        \caption{Clustering analysis of noiseless spectra with Fisher-EM. \textit{Top:} Convergence rate as a function of K. For every K value considered, 32 classifications were calculated. \textit{Bottom:} Boxplots of the ICL are a function of the number of clusters K. The horizontal bars show the median value, the boxes represent the two quartile values, the whiskers extend to points that lie within 1.5 times the interquartile range of the lower and upper quartile, and data beyond are shown individually with dots.}
        \label{ICL_SNRINF}
\end{figure}

\begin{figure}
        \centering
        \hfill
        \begin{subfigure}{0.5\textwidth}
            \centering
            \includegraphics[width=\hsize]{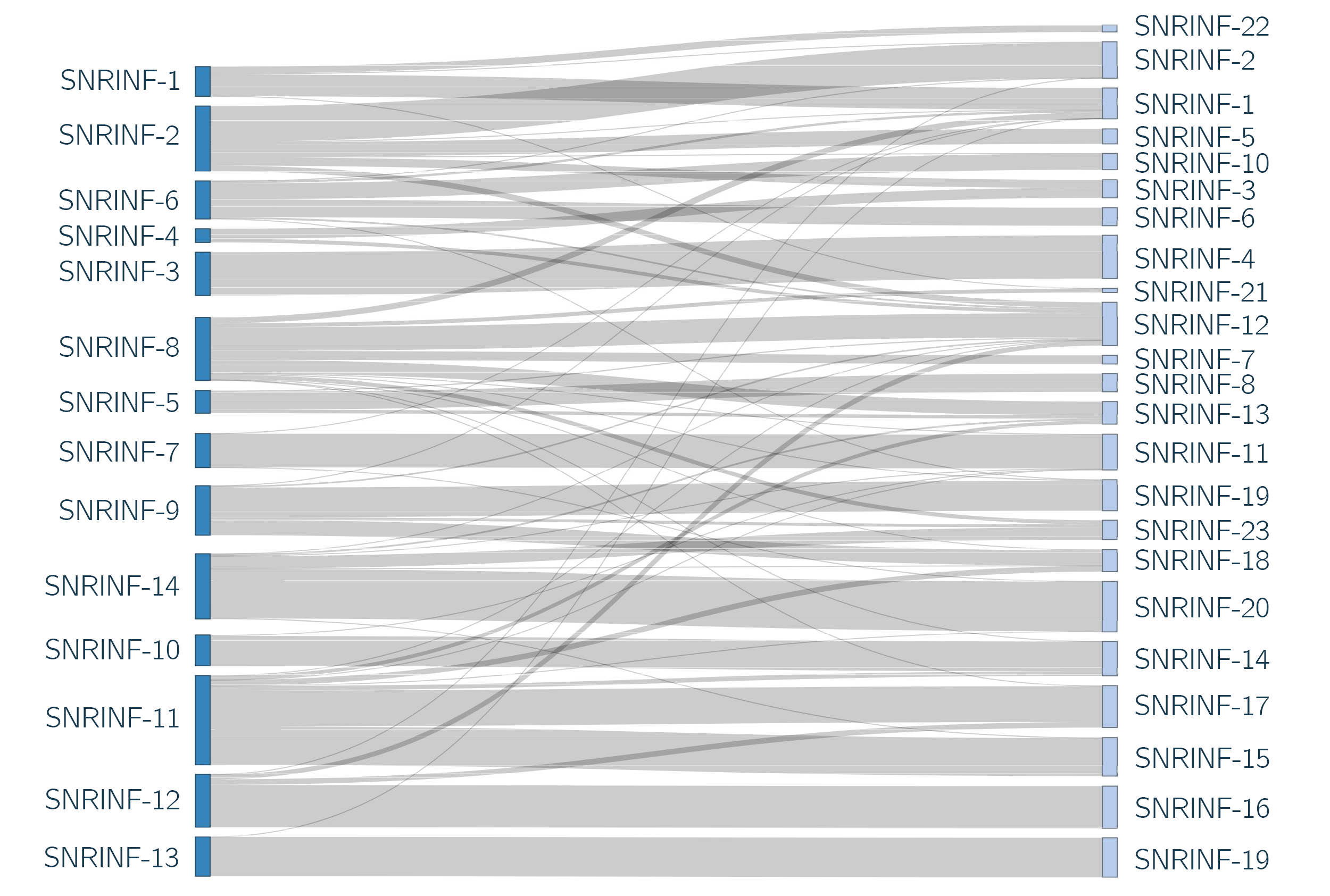}
            \caption{\textit{Left:}K=14. \textit{Right:} K=23}
            \label{fig:sankey_SNRINFK14K23}
        \end{subfigure}
        \hfill
        \begin{subfigure}{0.5\textwidth}
            \centering
            \includegraphics[width=0.95\hsize]{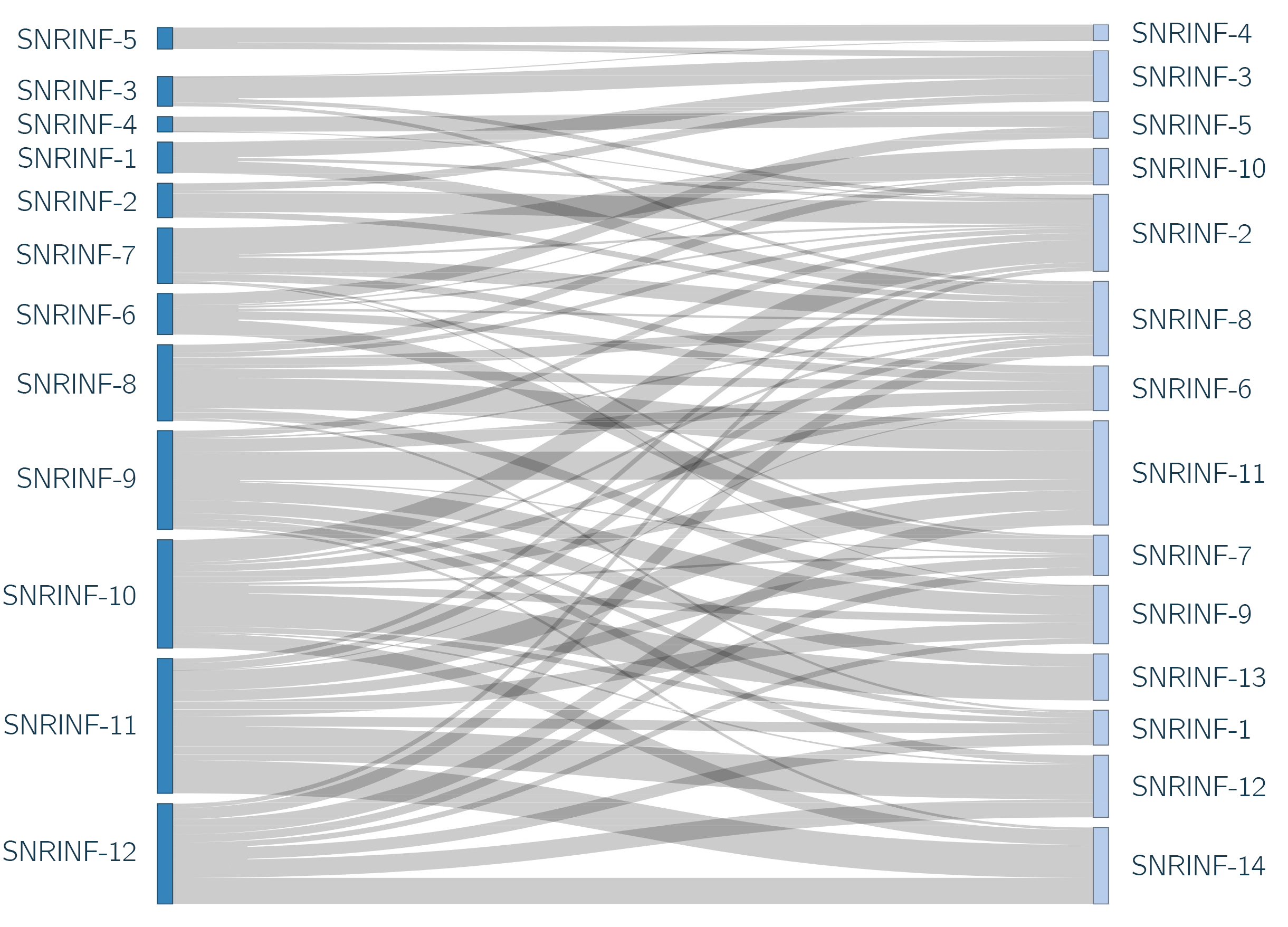}
            \caption{\textit{Left:}K=12. \textit{Right:} K=14}
            \label{fig:sankey_SNRINFK12K14}
        \end{subfigure}
        \caption{Comparisons between three classifications of the noiseless spectra (SNRINF): K=12, K=14, and K=23. In panel (a), the composition of the classes of K=14 (left) and K=23 (right) are compared, and K=12 (left) and K=14 (right) are compared in panel (b). The colour boxes represent the classes, the grey lines represent galaxies that are shared by the two classes they link, and the height of the colour boxes is proportional to the number of galaxies in a given class.}
        \label{fig:sankeySNRINF}
\end{figure}

In this section, we apply the Fisher-EM algorithm on the simulated noiseless spectra described in Sect. \ref{section:data}. 
There is a remarkable jump in ICL values at $K=14$ (Fig.~\ref{ICL_SNRINF}). In addition, the algorithm does not converge for $K=13,$ while there is a $100\%$ convergence rate at $K=12$ and $K=14$. Henceforth, $K=13$ appears as a frontier between two very distinct regimes. For $K<13$, the algorithm always converges, and the ICL is maximum at K=12. In the $K>13$ regime, the convergence rate decreases as the number of clusters increases, and no convergence is obtained for any $K>23$. Strictly speaking, the ICL is the greatest for $K=23,$ and from a statistical standpoint, $K=23$ is therefore the best result, but has a low convergence rate. In addition, the classification at $K=14$ has no convergence issue and matches the classification at $K=23$ very well in that they share many classes containing the same galaxies (i.e. they have a similar class composition, see Fig.~\ref{fig:sankey_SNRINFK14K23}). Undeniably, that the K=23 classification is more refined through its 9 additional clusters and a slightly better ICL. Nonetheless, the $K=14$ classification seems to be a good compromise between reproducibility  and goodness-of-fit, and it was therefore chosen to represent the $K>13$ regime. 

The classifications at $K=14$ and $K=12$ are arranged in a significantly different way (Fig.~\ref{fig:sankey_SNRINFK12K14}), showing that the two regimes in fact correspond to two distinct classifications.

\begin{figure}
    \centering
    \includegraphics[width=\hsize]{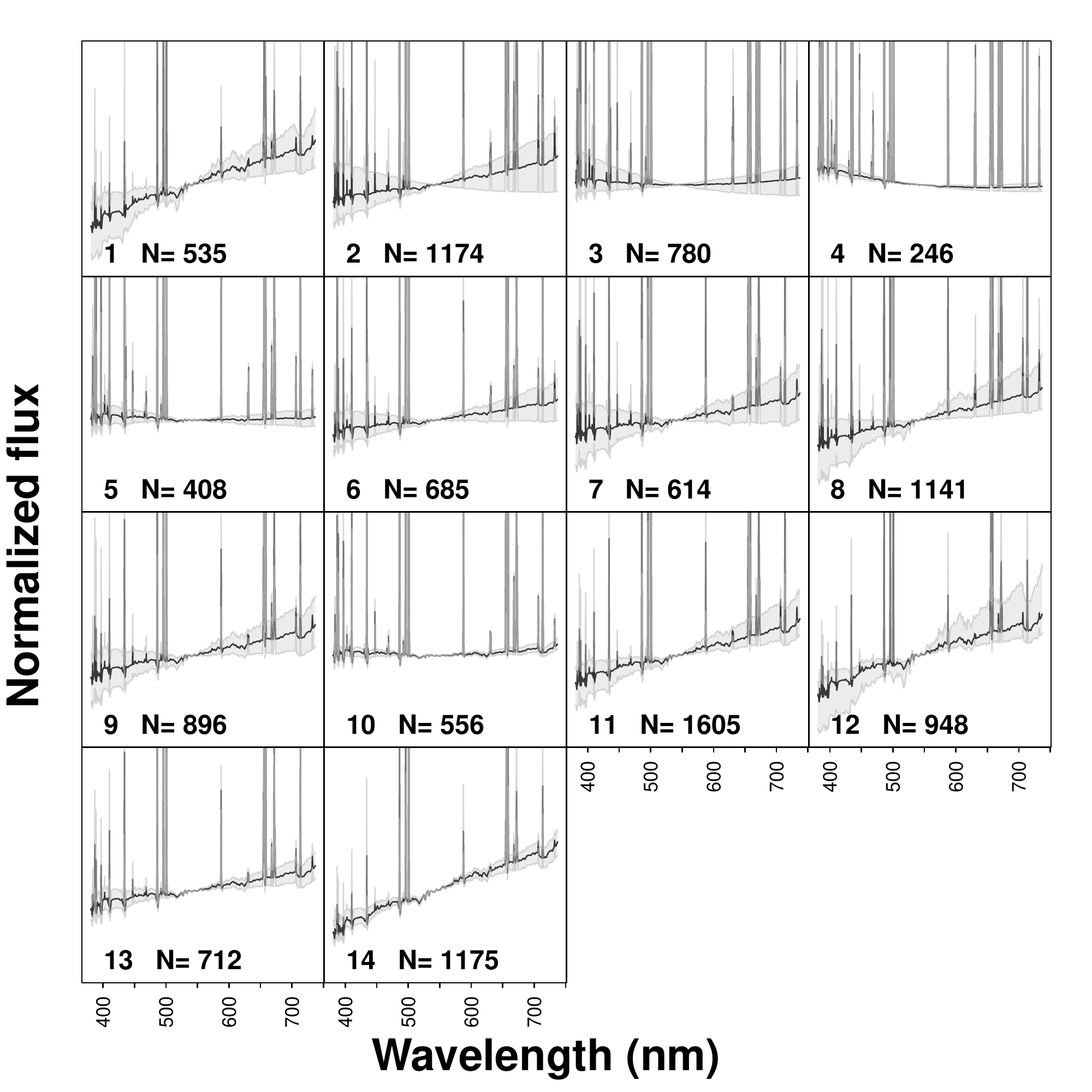}
    \caption{Fourteen-cluster classification of the noiseless spectra, with the mean spectra (in black) and their dispersion (in grey) for every class. N is the number of members in each class. All the spectra were normalised by their mean values between 505 and 581~nm, a region where the spectra have no emission lines. The scale is the same for all panels and all other figures of spectra throughout this paper. The dispersion corresponds to the 10\% and 90\% quantiles for each monochromatic flux. The classes are sorted by ascending average $T_{main}$.}
    \label{fig:spectra_SNRINFK14}
\end{figure}

\begin{figure}
    \centering
    \includegraphics[width=\hsize]{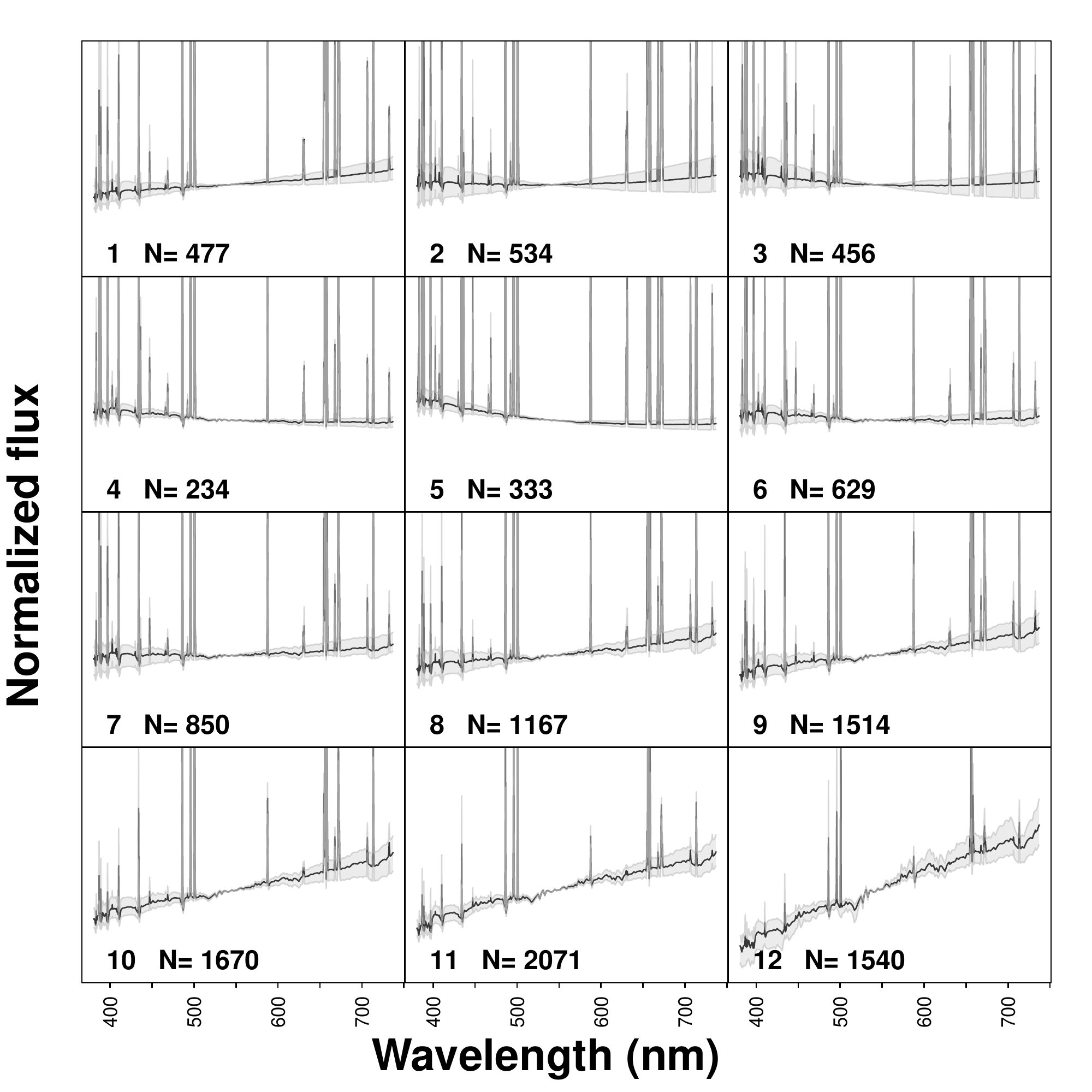}
    \caption{Twelve-cluster classification of the noiseless spectra (see Fig.~\ref{fig:spectra_SNRINFK14}). 
    }
    \label{fig:spectra_SNRINFK12}
\end{figure}

The mean spectra and dispersion of each class of the retained classifications (i.e. $K=14$ and $K=12$) are shown in Fig.~\ref{fig:spectra_SNRINFK14} and Fig.~\ref{fig:spectra_SNRINFK12} (the vertical scale is arbitrary, but it is the same for all plots of spectra). Despite its better ICL, the $K=14$ classification shows a higher dispersion than $K=12$ for most of the classes, except for a few exceptions (classes 4, 10, and 14), and in some cases, blue and red continua are even mixed (classes 2, 3, 6, and 9). On the other hand, the dispersion within most classes of the $K=12$ classification is rather small, indicating a decent homogeneity of the classes.

The distribution of the spectra among the 14 and 12 classes is relatively well balanced (Figs.~\ref{fig:heatmaps_SNRINFK14} and \ref{fig:heatmaps_SNRINFK12}), although this is arguably more the case even for the 14 classes. It varies from about 200 to 2000 spectra with an average of circa 500.

In both cases, there is no mean spectrum without emission lines. This is because several spectra with intense lines are present in all classes.
In particular, in our sample, emission lines are present in spectra with f$_{burst}$=0 that have a high $\tau_{main}$ (e.g. 5000). 

\subsection{Parameter distribution among classes}
\label{section:noiseless:pardistr}

\begin{figure}
    \centering
    \includegraphics[width=\hsize]{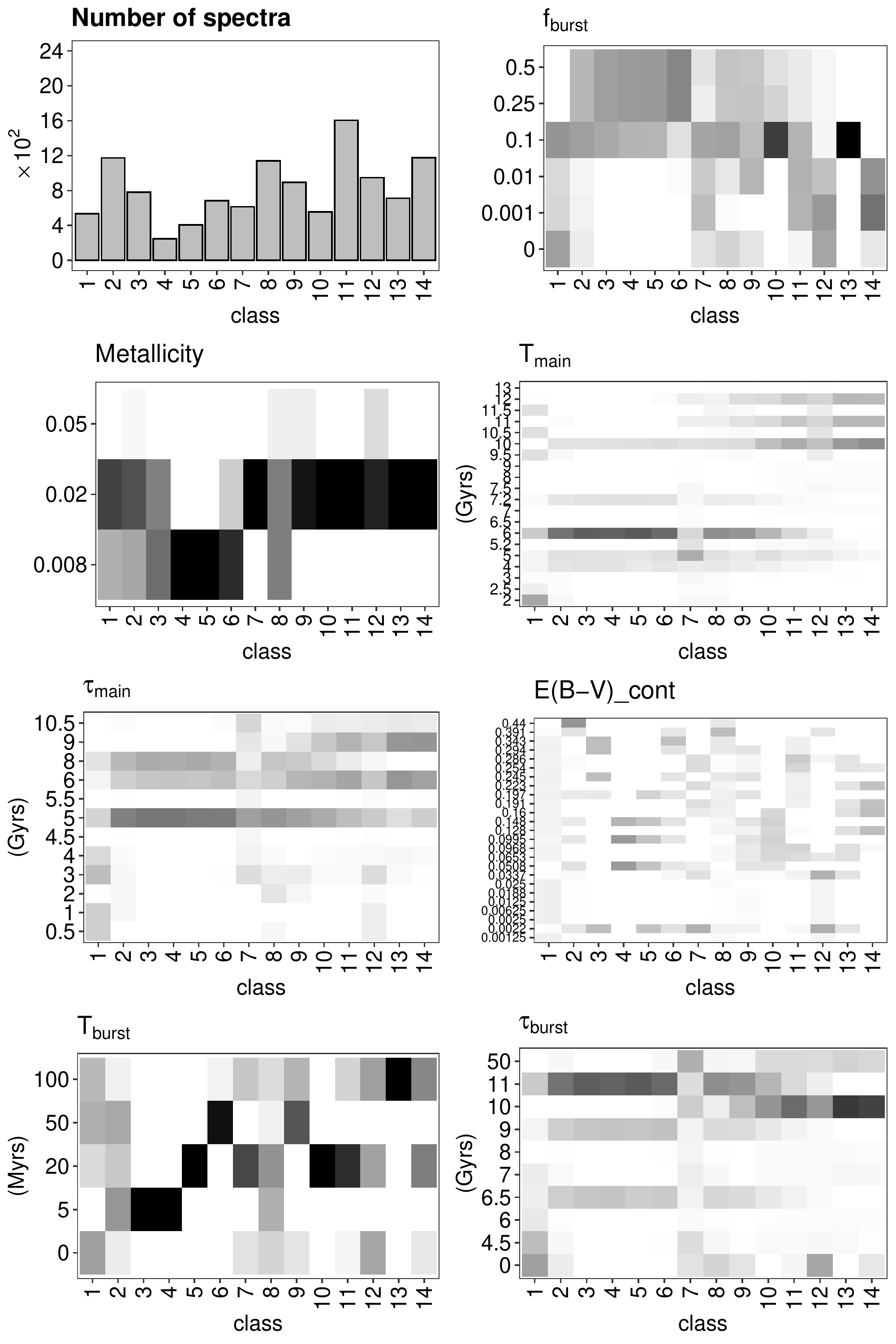}
    \caption{Fourteen-cluster classification of the noiseless spectra. \textit{Top left:} number of spectra contained in each class. \textit{All others:} heatmaps of the relevant CIGALE input parameters among the 14 classes on noiseless spectra. All possible parameter values (see Table~\ref{table:parameters}) are represented on the y-axis, and the class index on the x-axis. The within-class densities of the parameter values are illustrated in the form of a heatmap, where a dark square equates to a density of 1, and white of 0. The classes are sorted by ascending average T$_{main}$.}
    \label{fig:heatmaps_SNRINFK14}
\end{figure}

\begin{figure}
    \centering
    \includegraphics[width=\hsize]{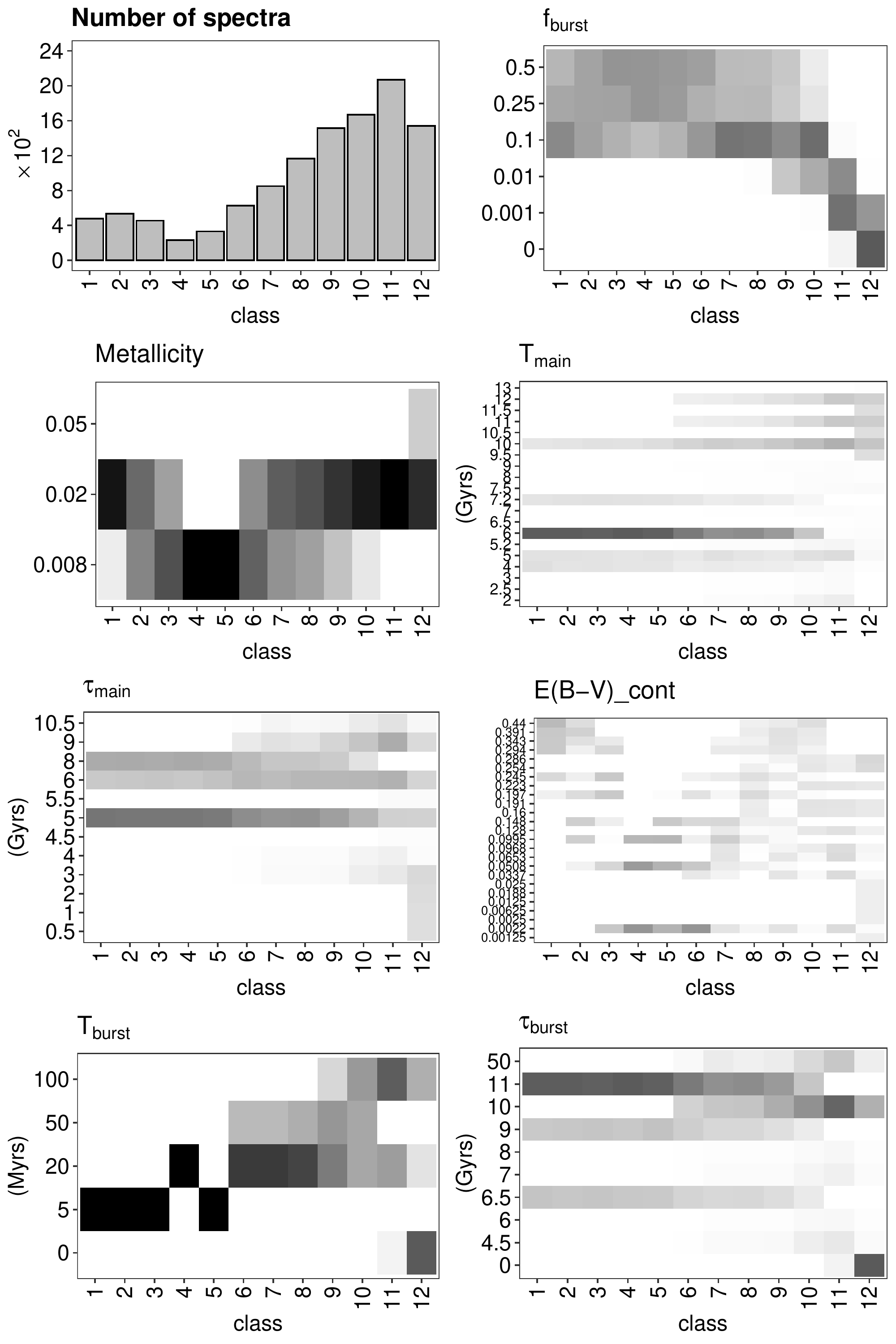}
    \caption{Twelve-cluster classification of the noiseless spectra (see Fig.~\ref{fig:heatmaps_SNRINFK14}).}
    \label{fig:heatmaps_SNRINFK12}
\end{figure}

As mentioned is section \ref{section:data}, the simulated spectra are associated each with a set of parameter values. As explained in Sect. \ref{section:data}, the relevant values are T$_{main}$, T$_{burst}$, $\tau_{main}$, $\tau_{burst}$, f$_{burst}$, metallicity, and E(B-V)$_{cont}$.

A parameter-by-parameter analysis and class-by-class analysis of the classifications is made possible by visualising the parameter distributions in the 12 and 14 classes (Fig.~\ref{fig:heatmaps_SNRINFK12} and Fig.~\ref{fig:heatmaps_SNRINFK14}).

\subsubsection{Classification at K=14} 
We first considered the parameter-by-parameter approach. f$_{burst}$ is mostly separated into two categories in this classification: higher values (classes 2, 3, 4, 5, 6, 10, and 13) and lower values (classes 1, 7, 8, 9, 11, 12, and 14). The separation is not very sharp, however, especially for some lower f$_{burst}$ classes.

The metallicity is not well sampled as it only takes three discrete values (0.008, 0.02, and 0.05), but is rather well segregated in the classes. Half of the classes gather one single metallicity value, two of them being the lower value (classes 4 and 5), and five being the medium value (classes 7, 10, 11, 13, and 14). The other half contain a mixture of usually two values: lower and medium (classes 1, 3, and 6) or medium and higher (classes 10 and 12), with the exception of classes 2 and 8, which mix all three values. The higher metallicity value is not as well separated as the medium and lower values.

The classes were sorted by ascending age T$_{main}$. A clear distinction is made in the classification between younger and older galaxies. Three categories can be drawn: classes containing the youngest galaxies (1 to 6), classes containing an equivalent mixture of young and old galaxies (7 to 12), and classes of older galaxies (13 and 14).

$\tau_{main}$ has a high dispersion in the classes, although the lower values happen to be concentrated in the classes with lower f$_{burst}$.

The T$_{burst}$ values are well isolated among the classes with high f$_{burst}$, with the exception of class 2, which also contains a fraction of low f$_{burst}$ galaxies. In the other classes, where the burst of star formation is less prominent, no distinction is made between T$_{burst}$ values.

$\tau_{burst}$ is quite dispersed in most of the classes with lower f$_{burst}$ classes. This is slightly less the case for higher f$_{burst}$ classes, but $\tau_{burst}$ remains a poorly discriminated parameter in the classification.

E(B-V)$_{cont}$ does not appear to be well separated. Except for a few classes (4, 10, and 14), high and low E(B-V)$_{cont}$ values are mixed in this classification.

All things considered, the classification at $K=14$ is essentially explained by four parameters (f$_{burst}$, T$_{main}$, T$_{burst}$, and metallicity), while the other three ($\tau_{main}$, $\tau_{burst}$, and E(B-V)$_{cont}$) show similar distributions for most of the classes, with a few exceptions. \\

The properties of the classes can be summarized as follows.
Classes 1 to 6 are made of galaxies of younger ages (T$_{main}$) and a rather significant burst of star formation (f$_{burst}$). Except for classes 1 and 2, these classes nearly only have one value of T$_{burst}$: Classes 3 and 4 contain galaxies whose bursts occurred very recently in the star formation history (5 Myr), class 6 contains galaxies of older bursts (50 to 100 Myr), and class 5 medium age bursts (20 Myr). Metallicity values are also fairly well separated in these classes. Galaxies of classes 2, 3, and 6 have metallicities of 0.008 and 0.02, while classes 4 and 5 only contain galaxies with a metallicity of 0.008.

Classes 7 to 12 mostly contain two populations of galaxies: older galaxies with a significant burst of star formation, and younger galaxies. Except for class 8, they all gather galaxies of medium to high metallicity. In this category, class 10 stands out as the galaxies it is made of all have a 20 Myr old prominent burst of star formation, while the other classes do not differentiate T$_{burst}$.

Finally, classes 13 and 14 gather old galaxies with a faint burst of star formation. More precisely, galaxies of class 14 have had little to no burst in their star formation history, while galaxies of class 13 did, but a long time prior to observation (old T$_{burst}$).

Overall, each class shows its own specificity in regard to the physics of the galaxies it contains. Three groups of classes can be distinguished (1-6, 7-12, and 13-14) which essentially categorise the galaxies as young and active, less active, and inactive and old.

\subsubsection{Classification at K=12}
\label{section:noiseless:params:K12}
As shown in Fig.~\ref{fig:sankey_SNRINFK12K14}, the classification at $K=12$ is significantly different from the classification at $K=14$ in terms of spectrum distribution. However, from a physical standpoint, they show very similar characteristics (Fig.~\ref{fig:heatmaps_SNRINFK12}).

They are both mostly driven by f$_{burst}$, metallicity, T$_{main}$ , and T$_{burst}$. Nonetheless, their distribution among the classes shows a significantly lower dispersion for $K=12$ despite the lower ICL. This is specifically striking for lower values of f$_{burst}$ , which are scattered around many classes and are mixed with higher values for $K=14$, while they are extremely well separated for $K=12$. 

Class-wise, a similar categorisation as in $K=14$ can be made, with classes 1-5 corresponding to the young active galaxies, classes 6-10 to a mixture of 
old active and younger galaxies, and classes 11-12 corresponding to the old and mostly inactive galaxies. In addition, class 12 gathers almost all galaxies that did not undergo an additional period of star formation, while that information was not retrieved at $K=14$.

As a whole, the classification at K=12 is more discriminative that the classification at K=14, but they are both sensitive to the same physical parameters. Despite the statistical superiority of K=14, K=12 therefore appears to be a better version of K=14 in terms of their physical discriminative properties.

\subsection{Linear discriminant analysis}
\label{section:noiseless:lda}

\begin{figure}
        \centering
        \includegraphics[width=\hsize]{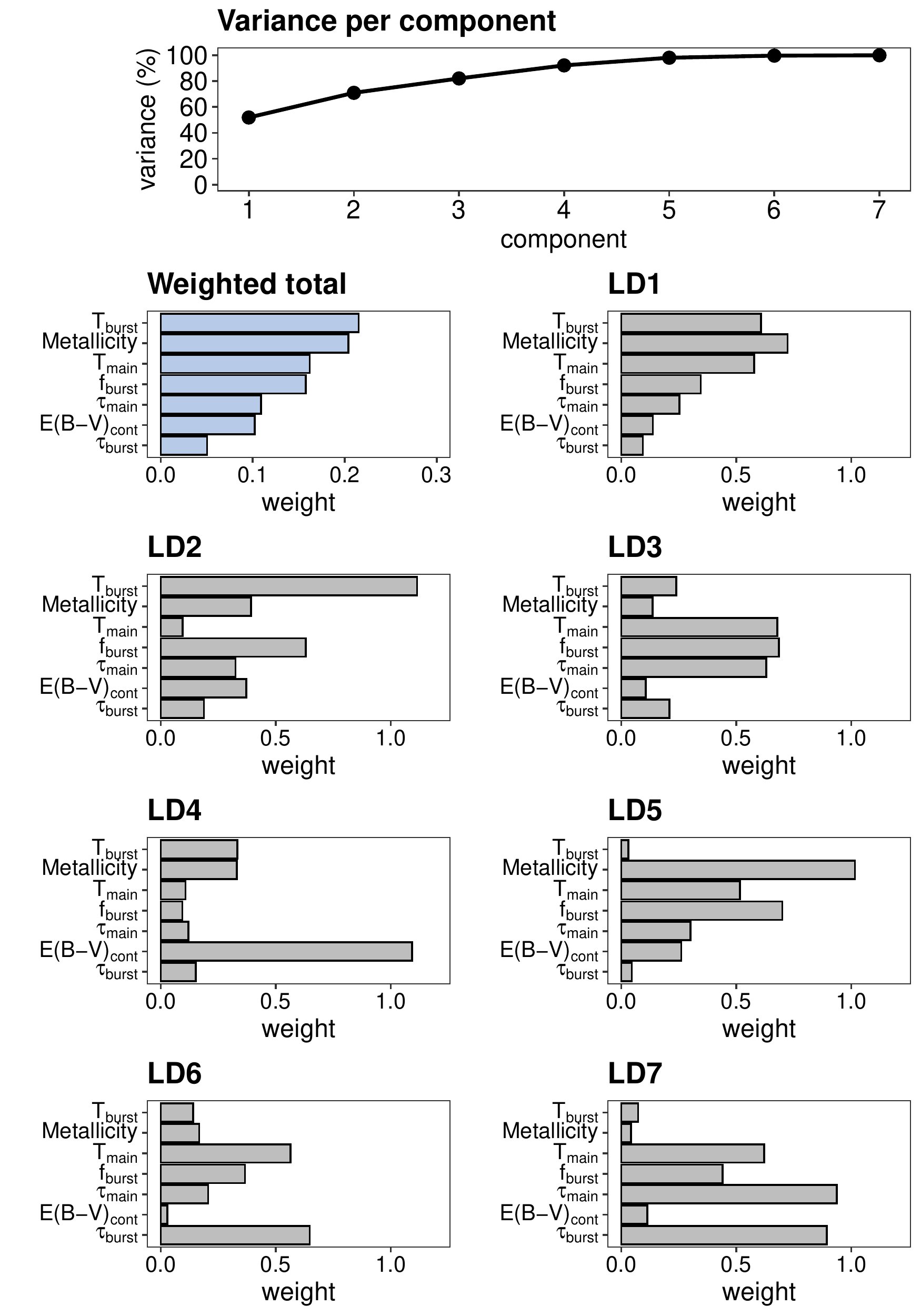}
        \caption{Linear discriminant analysis on the classification of noiseless spectra at $K=14$. \textit{Top:} Cumulative data variance described by the linear discriminant analysis components. \textit{LD1 to LD7:} Weight of each parameter for components 1 to 7 of the linear discriminant analysis. \textit{Weighted total:} Cumulative weight of each parameter among the seven components weighted by the percentage of data variance described by each component.}
        \label{fig:LDA_SNRINFK14}
\end{figure}

Linear discriminant analysis (LDA) was applied to the classified data in order to identify the influence of each parameter in the classification process. The LDA analysis returns a set of components that are essentially the projection vectors that best separates the classes, and while the classification was made based on the information of the spectra, LDA analysis uses the information of the parameters, therefore highlighting the links between them and the classification. The results are shown in Fig.~\ref{fig:LDA_SNRINFK14} and Fig.~\ref{fig:LDA_SNRINFK12}. 

\subsubsection{Classification at K=14}
The analysis at $K=14$ resulted in seven components labelled LD1 to LD7 (Fig.~\ref{fig:LDA_SNRINFK14}). The first compontent explains almost half of the data variance, and adding the next three brings it up to almost 90\%. 

An overall weight attributed to each parameter by the LDA is obtained by summing the seven components weighted by their relative variance explained. This overall weight quantifies how discriminated the parameters are by the classification. The results agree well with the conclusion obtained in Sect.~\ref{section:noiseless:pardistr}, namely, the parameters that are best discriminated are T$_{burst}$, metallicity, T$_{main}$ , and f$_{burst}$. The e-folding time of the burst of star formation is by far the least discriminated parameter, followed by the reddening and the e-folding time of the main stellar population.

The first component allocates most of its weight equally to T$_{burst}$, metallicity, and T$_{main}$. The second component is dominated by the age of the burst of star formation (T$_{burst}$) and to a lesser extent, by the stellar mass fraction of the burst (f$_{burst}$). The third component equally distributes most of its weight to three parameters, namely T$_{main}$, $\tau_{main}$ , and f$_{burst}$. The fourth component is completely dominated by the reddening (E(B-V)$_{cont}$), which was mostly ignored by the previous component. The final three components are mostly dominated by parameters that are already greatly taken into account in the first three components, but they also allocate some of their weight to the e-folding times of the SFR of the main stellar population and the burst event ($\tau_{main}$, $\tau_{burst}$), which were entirely insignificant in the previous components. 

\subsubsection{Classification at K=12}

\begin{figure}
        \centering
        \includegraphics[width=\hsize]{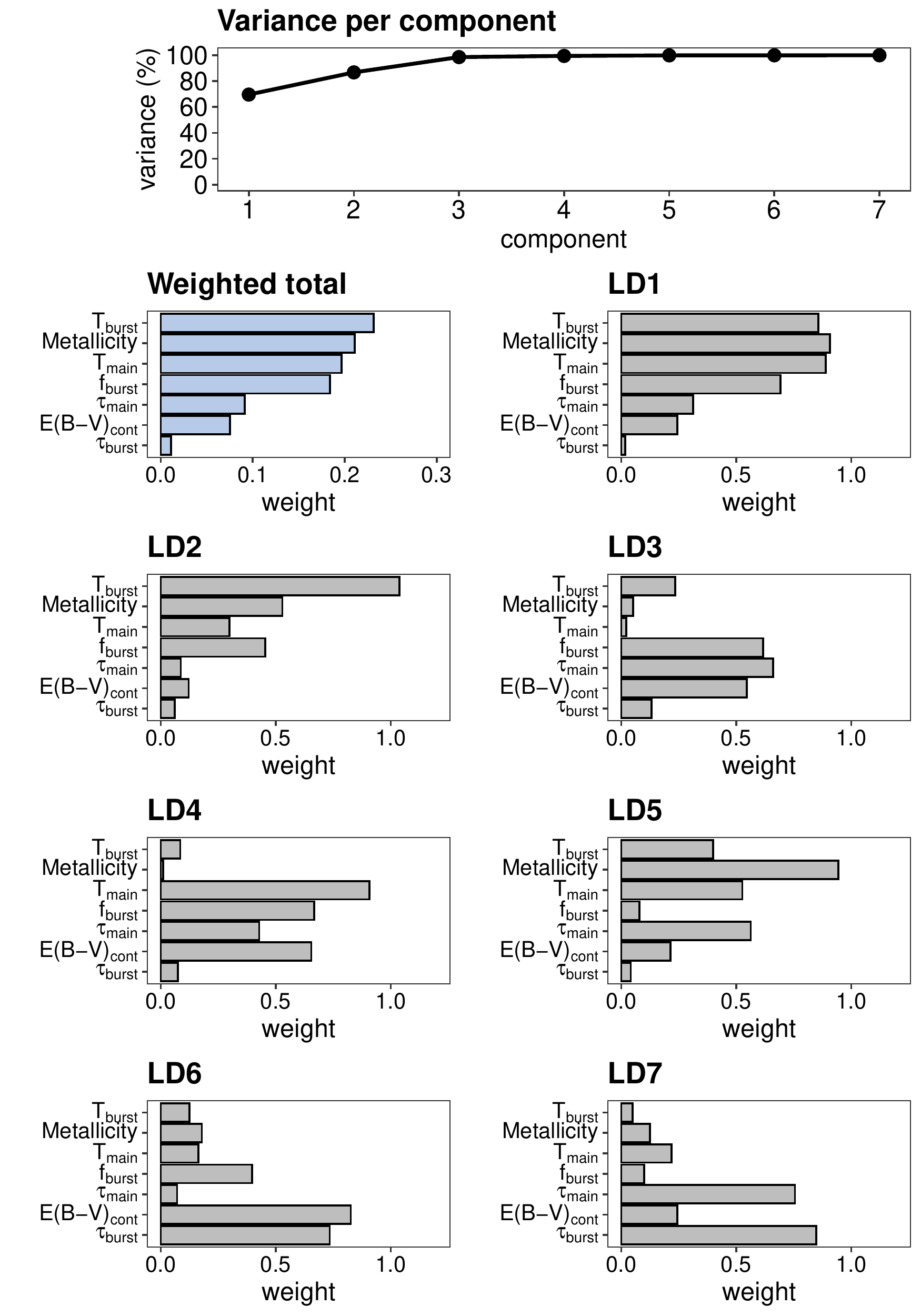}
        \caption{Linear discriminant analysis on the classification of noiseless spectra at $K=12$ (see Fig.~\ref{fig:LDA_SNRINFK14}).}
        \label{fig:LDA_SNRINFK12}
\end{figure}

As expected given their similarities, the LDA analysis at $K=12$ shows similar results than $K=14$ (Fig.~\ref{fig:LDA_SNRINFK12}). There is a significant difference between LD4 and LD6, however. Because the explained variance of those components is so small, however, they have little to no impact on the overall weight of the parameters.

\section{Analysis of the noisy spectra}
\label{section:noise}

In this section, we study the effect of noise on the classification of the spectra for different values of S/N. To do this, a Gaussian noise of constant S/N was added to the 11 475 spectra. We used seven of values for the S/N: 1, 3, 5, 10, 20, 100, and 500.

\subsection{Optimal number of clusters}
\label{section:noise:nbclusters}

\begin{figure}
        \centering
        \includegraphics[width=\hsize]{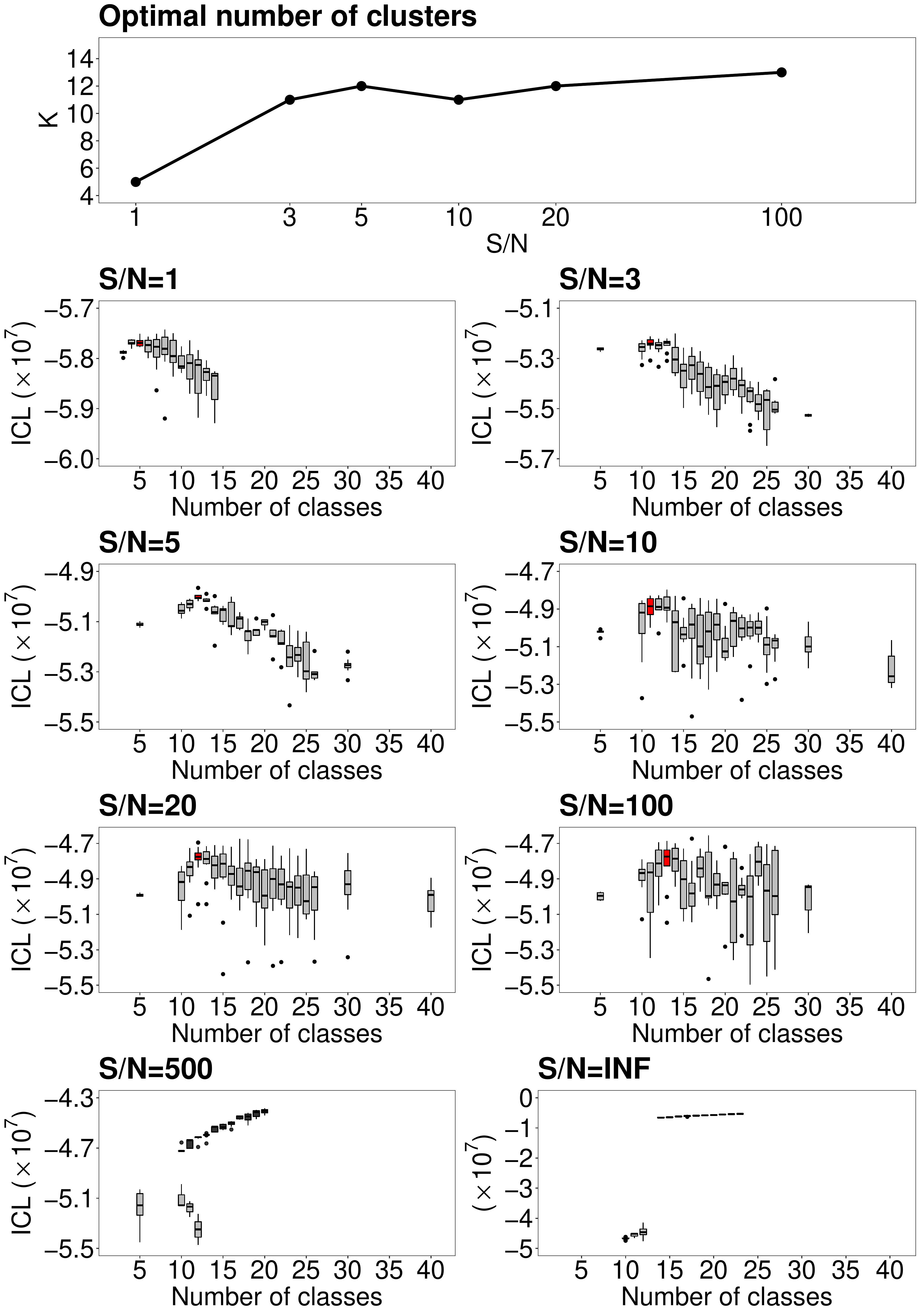}
        \caption{Summary of ICLs and optimal number of clusters as a function of S/N. \textit{Top:} Optimal number of clusters across the different noise levels. \textit{All others:} ICL as a function of K for different noise levels. In each of the panels, the red boxplot highlights the maximum median value of the ICL i.e. the associated best-fit K value. For the S/N of 500, two behaviours were observed depending on the randomly generated noise. They are both illustrated by the two sets of boxplots (black and grey) in the corresponding panel.}
        \label{ICL_noise}
\end{figure}

\begin{figure}
    \includegraphics[width=\hsize]{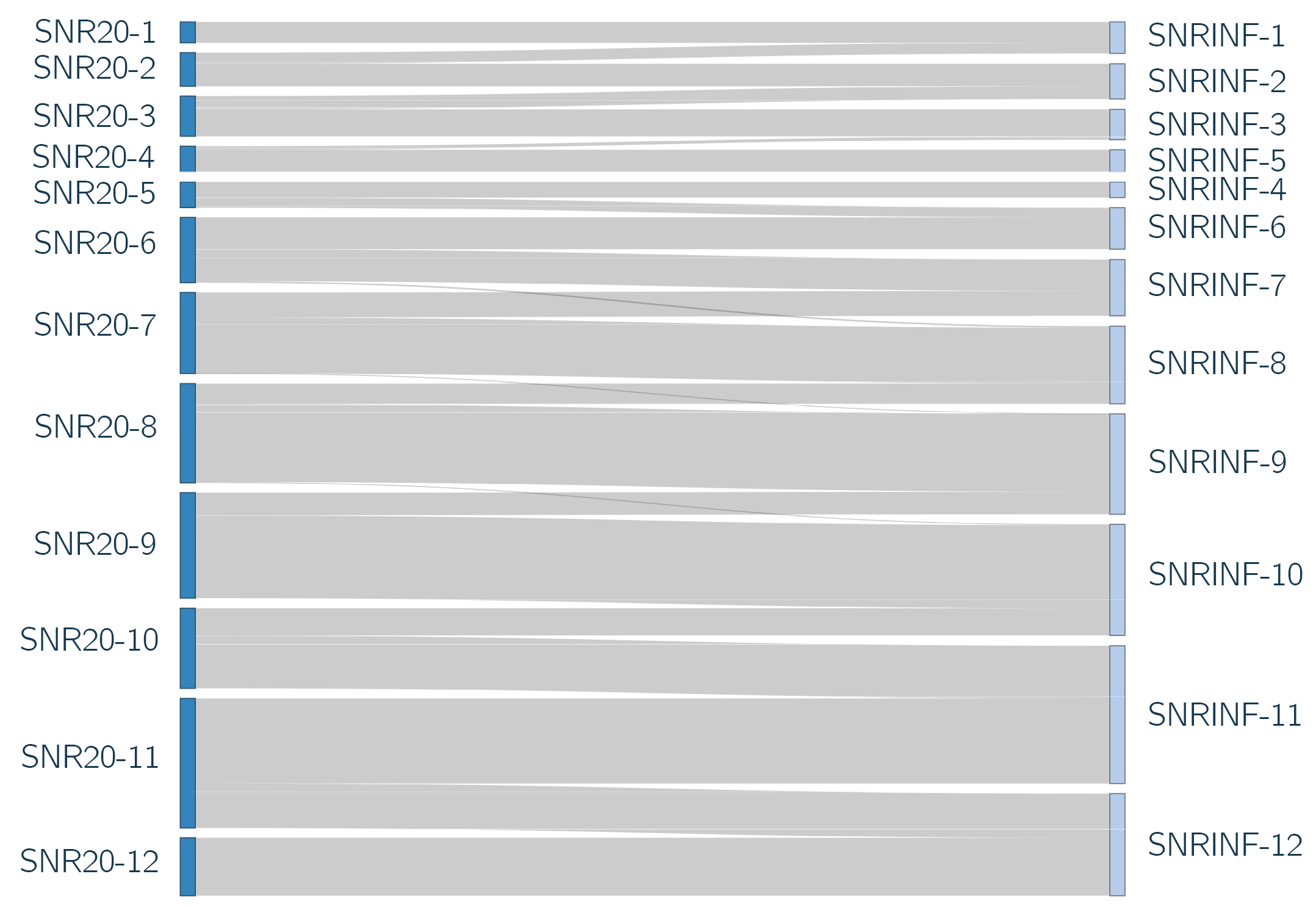}
    \caption{Same as Fig.~\ref{fig:sankeySNRINF} between the $K=12$ classification on noiseless spectra (right) and on spectra with an added noise of S/N=20 (left).}
    \label{fig:sankey_SNR20K12_SNRINFK12}
\end{figure}

\begin{figure}
    \includegraphics[width=\hsize]{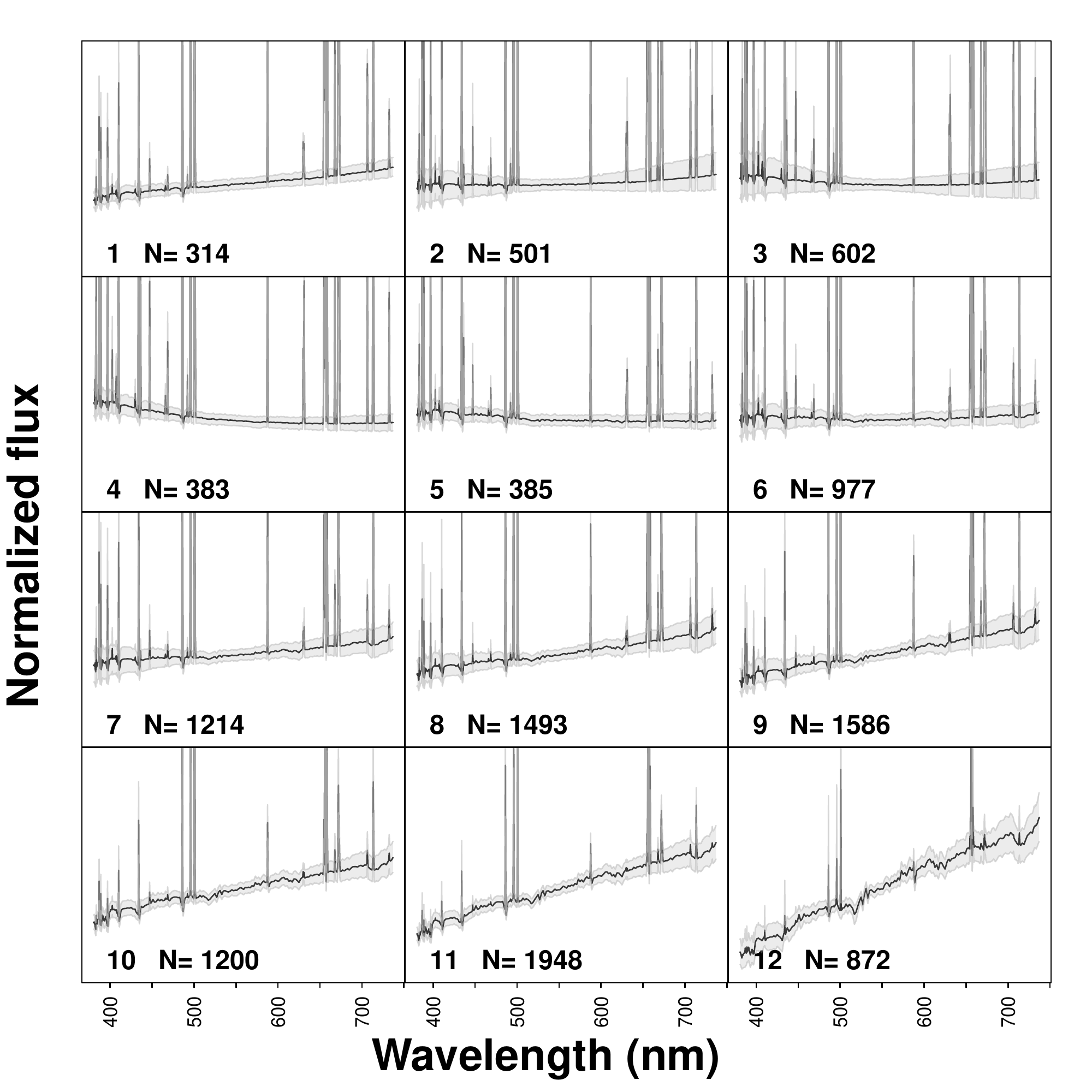}
    \caption{Twelve-cluster classification obtained on the spectra with added noise of S/N=20 (see Fig.~\ref{fig:spectra_SNRINFK14}).}
    \label{fig:spectra_SNR20K12}
\end{figure}

The very characteristic break of the ICL curve observed at K=13 for the noiseless spectra disappears as soon as noise as low as S/N=500 is added. The ICL curves obtained have an optimum independently of the noise level (Fig.~\ref{ICL_noise}), as opposed to the ever-increasing ICL on noiseless spectra. 

For S/N$\leq$100 and below, the ICL reaches its maximum for K=11 to 13. In addition, convergence is reached every single time for any K smaller than 30-40, depending on the noise level. At S/N=500, the ICL sometimes shows another behaviour that fully depends on the random generation of the noise. 
At S/N=500, the noise may be insufficient to blur out the data sparsity, which is likely responsible for the ICL break and convergence issues on noiseless spectra. Therefore, while for most generated noise the ICL showed an optimum, one particular noise vector led to an ever-increasing ICL curve (until loss of convergence) that resembles that of the noiseless spectra.
    
Our study of the noise shows that the ICL curves at different S/N differ from that of the noiseless case, but they all agree and yield the same optimal number of clusters around K=12. Furthermore, we show that the optimal classifications on the spectra with added noise closely resembles the $K=12$ one on noiseless spectra, whether it be based on the composition of the classes (Fig.~\ref{fig:sankey_SNR20K12_SNRINFK12}) or their median spectrum (Fig.~\ref{fig:spectra_SNR20K12}).

The classification at S/N=20 and K=12 is taken as a reference in the rest of this paper. The results for the other S/N are available in Appendix~\ref{appendix:noise}. 

\subsection{Parameter distribution among classes}
\label{section:noise:pardistr}

\begin{figure}[t]
        \centering
        \includegraphics[width=\hsize]{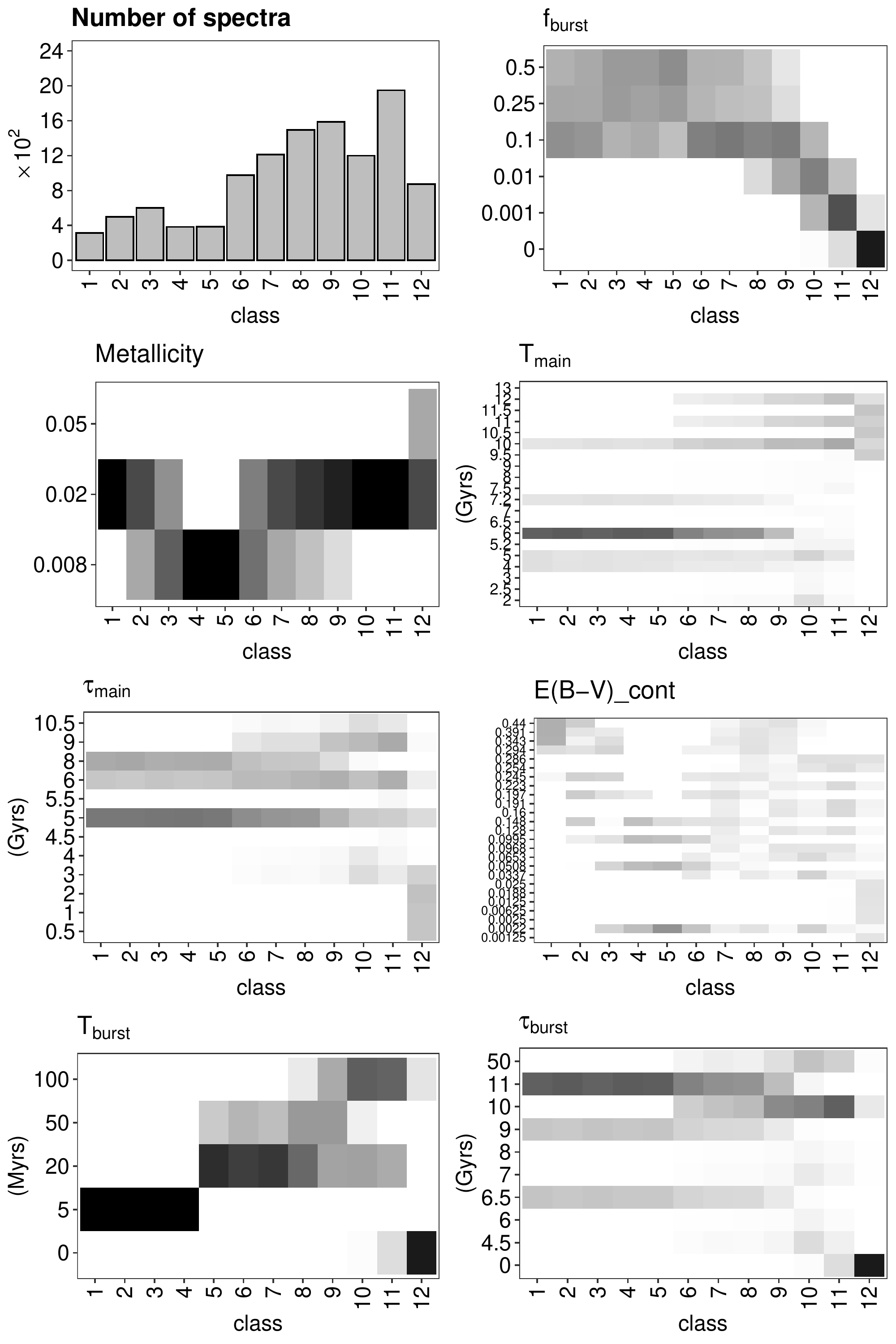}
        \caption{Twelve-cluster classification on the spectra with an added noise of S/N=20 (see Fig.~\ref{fig:heatmaps_SNRINFK14}).
        }
        \label{figure:heatmaps_SNR20K12}
\end{figure}

The optimal classification at S/N=20 is essentially identical to that on the noiseless spectra presented in Sect.~\ref{section:noiseless:params:K12} in regard to the parameter distribution in the classes (Fig.~\ref{figure:heatmaps_SNR20K12}). Slight differences appear nonetheless, highlighting the loss of information induced by the additional noise. For example, there is no longer a class isolating 20 Myr old star bursts. On the other hand, some specificities in the spectra appear to be retrieved more accurately with the addition of noise. For instance, five classes contain a unique value of metallicity, as opposed to three classes on noiseless spectra. Lower values of f$_{burst}$ are also more sharply separated. 

At greater noise (Appendix~\ref{appendix:noise}), this enhanced ability to discriminate some parameters fades out, and the dispersion in the classes increases for all parameters. At S/N=3, the method is still capable of distinguishing burstless galaxies from burst-heavy ones, but lower non-zero values of f$_{burst}$ become more erratically distributed around the classes. Metallicity is not as well discriminated either, as only a single class of unique value remains. T$_{main}$, and T$_{burst}$ to a lesser extent, is still separated in a similar fashion despite the significant amount of noise.
At S/N=1, the optimum shifts from 12 to 5 classes, but the method is still capable of approximately separating old and inactive galaxies from young and active ones.

\subsection{Linear discriminant analysis}
\label{section:noise:lda}
The LDA applied on the noisy spectra with S/N=20 shows that the first three components completely dominate the analysis (60\%, 20\%, and 10\%), whereas five components were significant for the noiseless spectra. The first component is similar to that of the noiseless spectra, with a more heavily weighted f$_{burst}$ parameter nonetheless. The second and third components are distinctively different, however. It appears that the weight of the parameters was shifted from one component to another: T$_{burst}$, f$_{burst}$ , and $\tau_{main}$ are less significant in the second component, but more important in the third component. Likewise, T$_{main}$, $\tau_{main}$ , and E(B-V)$_{cont}$ have higher weights in the second component and lower weight in the third component. While some components are indeed different, the overall relevance of each parameter in regard to the classification remains almost unchanged. 

\begin{figure}
        \centering
        \includegraphics[width=\hsize]{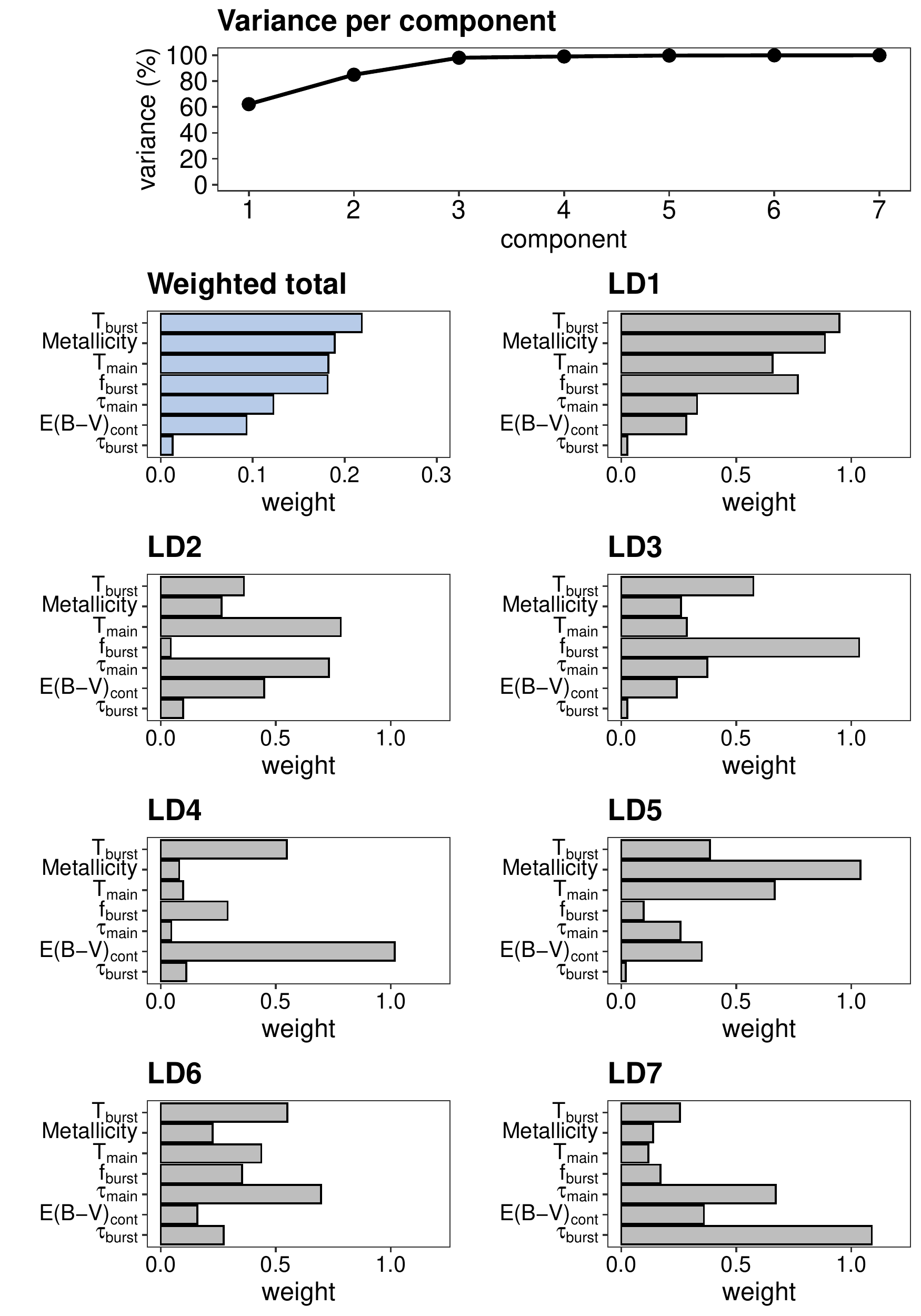}
        \caption{Linear discriminant analysis on the classification of spectra with added noise of S/N=20 (see Fig.~\ref{fig:LDA_SNRINFK14}).}
        \label{figure:LDA_SNR20}
\end{figure}

\section{Discussion}
\label{section:discussion}

\subsection{Origin of the K$\geq$13 regime}
\label{section:discussion:sparsity}

The analysis of the noiseless spectra revealed an odd behaviour at K$\geq$13: a lack of convergence at K=13, and a high plateau of the ICL for K from 14 to 23 (Fig.~\ref{ICL_SNRINF}). In an ideal world, the ICL curve as a function of the number K of clusters should show a clear peak because this criterion takes into account the number of free parameters in the statistical models and the good separation of the clusters. The ICL curve does not always have this ideal behaviour, however, especially in high dimensions \citep[e.g. ][]{Fraix-Burnet2021}. In particular, it often shows a plateau that ends when the algorithm fails to converge (i.e. encounters an empty cluster, see Sect.~\ref{section:noiseless:nbclusters}). 

Because the classifications for K$>$13 differ from the classifications obtained for K=12 with and without noise, we suspected that the reason might be that the data are simulated with a necessarily limited coverage of the parameter space. To test this hypothesis, we devised a toy model at the lowest possible dimension to reproduce this behaviour. After trials and errors, we found a dataset that is described in Appendix~\ref{toymodel}. The ICL curves show an optimum at K=4, with solutions at K=2 but no convergence at K=3. Remarkably, if we add a very small amount of noise to a part of one of the five variables, solutions are found in all cases, exactly as with our CIGALE data.

Even if this toy model cannot be considered as a proof, we conclude that the behaviour found in the noiseless data may be due to some peculiar distribution of the data in the parameter space, that is, too small a dispersion and probably a significant level of sparsity. In addition, \citet{Jouvin2021} reported a poor performance of the Fisher-EM results in the case of very little noise, a behaviour that they were unable to explain but hypothesised to be related to insufficient constraints brought by the dataset. Such problems are known to occur in EM-GMM-based clustering \citep[e.g.][]{Kasa2020}.

The case of K$\geq$13 is therefore thought to be an artefact resulting from the simulated nature of the spectra, and is dismissed in the rest of this section. Instead, K=12 is considered as the representative classification of the noiseless spectra.
We stress that the noiseless situation cannot be encountered in reality.

\subsection{Physical discrimination capacity of unsupervised classification}
\label{section:discussion:noiseless}

The classification at K=12 shows classes of spectra that are very homogeneous with little dispersion, demonstrating the ability of Fisher-EM to find structures in a high-dimensional data space. This has been noted before for the much larger SDSS sample \citep{Fraix-Burnet2021}. 

The analysis of the distribution of the parameters used in the CIGALE simulations shows that this discriminative power among the spectra is also visible in the physical properties of the galaxies. Four of the seven parameters are clearly well discriminated (T$_{main}$, T$_{burst}$, f$_{burst}$ , and metallicity), and to a lesser extent, $\tau_{main}$ and E(B-V)$_{cont}$. This important result shows that Fisher-EM is capable of picking up the expected relevant physical parameters. The LDA analysis confirms these most influential parameters. The fact that the weights of T$_{burst}$ and metallicity appear stronger than that of T$_{main}$ 
could be due to their small sampling density (three and four values, respectively). However, f$_{burst}$ has only six values and has a similar weight as T$_{main}$. Moreover, the latter has a higher weight than $\tau_{main}$ despite having a similar distribution. Lastly, the LDA analysis shows that $\tau_{burst}$ has a weak impact on the classification, but this is probably not due to its small sampling density since it was expected from the physics itself (Sect.~\ref{section:data:params}). 

As a conclusion, we have shown that each class not only has a specific spectral shape, but its members also have specific physical properties. Hence, in a real dataset, a detailed analysis of the mean spectra of the classes should reveal these properties and transform an unsupervised classification into an objective and physical atlas of galaxy spectra.

\subsection{Effect of the noise}
\label{section:discussion:noise}

The addition of noise raises two questions that we address in this section: i) whether it changes the classification itself, and, ii) whether it changes the physical interpretation.

\subsubsection{Effect on the classification}

Adding some noise to our spectra strongly modifies the ICL curve by revealing a clear maximum around K=12 (for S/N$\ge$3) with a quasi-identical classification, the noiseless one included. At S/N=1, the optimum is K=5 so that a higher level of noise tends to smear out the classes, and as expected, lessens the discriminative capability of the analysis.

The presence of noise in the data also tends to facilitate the convergence of Fisher-EM. At the S/N we considered, the convergence issues that were encountered in the noiseless case were non-existent. Lack of convergence was still observed in the noisy case when a high number of classes was chosen, but this behaviour is usual and was seen on real data as well \citep{Fraix-Burnet2021}.

\subsubsection{Effect on the physical meaning of the classes}

Our study shows that our unsupervised classification and its physical meaning are essentially unchanged between S/N=3 and S/N=500 for our simulated data, demonstrating its robustness.
At S/N=1, most of the physical discriminative capacity of the method is lost, and only five classes are found. Nonetheless, these five classes are not meaningless, and remarkably separate active from inactive galaxies rather well. \citet{Jouvin2021} showed that in some cases, Fisher-EM was even capable of accurately classifying data with a S/N as low as -1dB.

\section{Conclusion}
\label{section:conclusion}

This study shows that the unsupervised classification algorithm Fisher-EM applied on thousands of CIGALE galaxy spectra yields a classification that is both robust against the initialisation of the algorithm and against the noise. Very importantly, the classification is very discriminating with respect to the physical properties of the galaxies.

Unsupervised classification in astrophysics is still in its infancy, and the first robust classification of spectra of galaxies have been published very recently \citep{Fraix-Burnet2021}. The aim of such an objective classification is to produce an atlas that is entirely data driven and that could be used later with supervised learning in large surveys. Even though the preliminary interpretation of the classes found in \citet{Fraix-Burnet2021} has shown their physical relevance, we here confirmed that unsupervised machine learning is able to yield not only a robust statistical classification, but also a physical classification of the properties of the galaxies from the spectra. 

The main advantage of the unsupervised classification is that we do not add any a priori physical information into the classification process, but rely on the ability of the algorithm to detect the structures that it is really able to detect, not those we would wish it to detect. We are thus not limited by the representativeness of the training set, and we consequently avoid all the associated biases. In addition, as shown in \citet{Fraix-Burnet2021}, the classification is characterised by the statistical model that we can use for further supervised classifications. This is performed through the E-step only, which is extremely fast. This means that the Fisher-EM algorithm has built its own training set (Fig.~\ref{fig:spectra_SNRINFK12}) that happens to be physically well characterised (Fig.~\ref{fig:heatmaps_SNRINFK12}).

Our result is a strong encouragement to analyse the atlas proposed in \citet{Fraix-Burnet2021} in more depth and extend it to larger samples. The exciting perspective is to include galaxies at higher redshifts in order to study the evolution of the classification with time through a fully data-driven procedure.

\begin{acknowledgements}
We warmly thank Charles Bouveyron for many discussions during this study. This research has made use of the NASA/IPAC Extragalactic Database (NED), which is funded by the National Aeronautics and Space Administration and operated by the California Institute of Technology.
\end{acknowledgements}

\bibliographystyle{aa} 
\bibliography{cigale}

\begin{appendix}
\onecolumn
\section{Spectrum generation}

\begin{table*}[hbt!]
    \tiny
    \centering
    \caption{CIGALE parameters used to generate the data}
        \ra{1.3}
        \resizebox{0.7\textwidth}{!}{\begin{tabular}{|c|c|c|c|c|c|c|c|c|}
        \hline\hline
        \makecell{T$_{main}$\\(Myr)} & \makecell{$\tau_{main}$\\(Myr)} & f$_{burst}$ & \makecell{T$_{burst}$\\(Myr)} & \makecell{$\tau_{burst}$\\(Myr)} & metallicity & E\_BV\_lines & E\_BV\_factor & redshift\\
        \hline
        \multirow{5}{*}{\makecell{9500\\10500\\11500}} & \multirow{5}{*}{\makecell{500\\1000\\3000}} & \multirow{23}{*}{0} & \multirow{23}{*}{-} & \multirow{23}{*}{-} & \multirow{5}{*}{\makecell{0.02\\0.05}} & \multirow{5}{*}{\makecell{0.0005\\0.01\\0.025\\0.05\\0.075\\0.1}} & \multirow{5}{*}{0.25} & \multirow{5}{*}{\makecell{0.001\\0.002\\0.003\\0.004\\0.005}} \\
        & & & & & & & &\\
        & & & & & & & &\\
        & & & & & & & &\\
        & & & & & & & &\\ 
        \cline{1-2}\cline{6-9}
        \multirow{8}{*}{\makecell{10000\\11000\\12000}} & \multirow{8}{*}{\makecell{2000\\5000}} &  &  &  & \multirow{35}{*}{0.02} & \multirow{35}{*}{\makecell{0.005\\0.077\\0.148\\0.220\\0.292\\0.363\\0.435\\0.507\\0.578\\0.650}} & \multirow{41}{*}{0.44} & \multirow{8}{*}{\makecell{0.01\\0.02\\0.03\\0.04\\0.05\\0.06\\0.07\\0.08\\0.09}}\\
        & & & & & & & &\\
        & & & & & & & &\\
        & & & & & & & &\\
        & & & & & & & &\\
        & & & & & & & &\\
        & & & & & & & &\\
        & & & & & & & &\\
        \cline{1-2}\cline{9-9}
        \multirow{3}{*}{\makecell{6000\\10000}} & \multirow{3}{*}{5000} & & & & & & & \multirow{3}{*}{\makecell{0.1\\0.2\\0.3}}\\
        & & & & & & & &\\
        & & & & & & & &\\\cline{1-2}\cline{9-9}
        \multirow{3}{*}{\makecell{5000\\7500}} & \multirow{3}{*}{5500} & & & & & & & \multirow{3}{*}{\makecell{0.4\\0.5\\0.6}}\\
        & & & & & & & &\\
        & & & & & & & &\\\cline{1-2}\cline{9-9}
        \multirow{2}{*}{\makecell{4000\\6500}} & \multirow{2}{*}{4500} & & & & & & &\multirow{2}{*}{\makecell{0.7\\0.8}}\\
        & & & & & & & & \\\cline{1-2}\cline{9-9}
        13000 & 5000 & & & & & & & 0\\
        \cline{1-5}\cline{9-9}
        \multirow{3}{*}{\makecell{2000\\5200}} & \multirow{3}{*}{3000} & \multirow{18}{*}{\makecell{0.001\\0.01\\0.1}} & \multirow{18}{*}{\makecell{20\\100}} & \multirow{3}{*}{4500} & & & & \multirow{3}{*}{\makecell{0.8\\0.9\\1}}\\
        & & & & & & & &\\
        & & & & & & & &\\\cline{1-2}\cline{5-5}\cline{9-9}
        2000 & \multirow{4}{*}{4000} & & & \multirow{2}{*}{6000} & & & & \multirow{2}{*}{0.7}\\
        7000 & & & & & & & &\\\cline{1-1}\cline{5-5}\cline{9-9}
        2500 & & & & \multirow{2}{*}{7000} & & & & \multirow{2}{*}{0.6}\\
        7500 & & & & & & & &\\\cline{1-2}\cline{5-5}\cline{9-9}
        3000 & \multirow{4}{*}{5000} & & & \multirow{2}{*}{7000} & & & & \multirow{2}{*}{0.5}\\
        8000 & & & & & & & & \\\cline{1-1}\cline{5-5}\cline{9-9}
        4000 & & & & \multirow{2}{*}{8000} & & & & \multirow{2}{*}{0.4}\\
        9000 & & & & & & & &\\\cline{1-2}\cline{5-5}\cline{9-9}
        \multirow{3}{*}{\makecell{5000\\10000}} & \multirow{3}{*}{\makecell{5000\\10500}} & & & \multirow{3}{*}{50000} & & & & \multirow{3}{*}{\makecell{0.1\\0.2\\0.3}}\\
        & & & & & & & &\\
        & & & & & & & &\\\cline{1-2}\cline{5-5}\cline{9-9}
        \multirow{4}{*}{\makecell{10000\\11000\\12000}} & \multirow{4}{*}{\makecell{6000\\9000}} & & & \multirow{4}{*}{\makecell{10000}} & & & & \multirow{8}{*}{\makecell{0.01\\0.02\\0.03\\0.04\\0.05\\0.06\\0.07\\0.08\\0.09}}\\
        & & & & & & & &\\
        & & & & & & & &\\
        & & & & & & & &\\
        \cline{1-7}
        \multirow{4}{*}{6000} & \multirow{4}{*}{\makecell{5000\\8000}} & \multirow{10}{*}{\makecell{0.1\\0.25\\0.5}} & \multirow{10}{*}{\makecell{5\\20\\50}} &\multirow{4}{*}{11000} & \multirow{10}{*}{\makecell{0.008\\0.02}} & \multirow{10}{*}{\makecell{0.005\\0.116\\0.226\\0.337\\0.447\\0.558\\0.668\\0.779\\0.889\\1.000}} & &\\
        & & & & & & & &\\
        & & & & & & & &\\
        & & & & & & & &\\\cline{1-2}\cline{5-5}\cline{9-9}
        \multirow{3}{*}{\makecell{4000\\7200}} & \multirow{3}{*}{5000} & & & \multirow{3}{*}{6500} & & & & \multirow{3}{*}{\makecell{0.4\\0.5\\0.6}}\\
        & & & & & & & &\\
        & & & & & & & &\\
        \cline{1-2}\cline{5-5}\cline{9-9}
        \multirow{3}{*}{\makecell{5000\\10000}} & \multirow{3}{*}{6000} & & & \multirow{3}{*}{9000} & & & & \multirow{3}{*}{\makecell{0.1\\0.2\\0.3}}\\
        & & & & & & & &\\
        & & & & & & & &\\
        \cline{1-9}
        \hline
    \end{tabular}}
\label{table:parameters}
\end{table*}

\clearpage
\twocolumn
\section{Results for spectra with added noise}
\label{appendix:noise}

\subsection{Parameter distribution}
\begin{figure}[h!]
        \centering
        \includegraphics[scale=0.34]{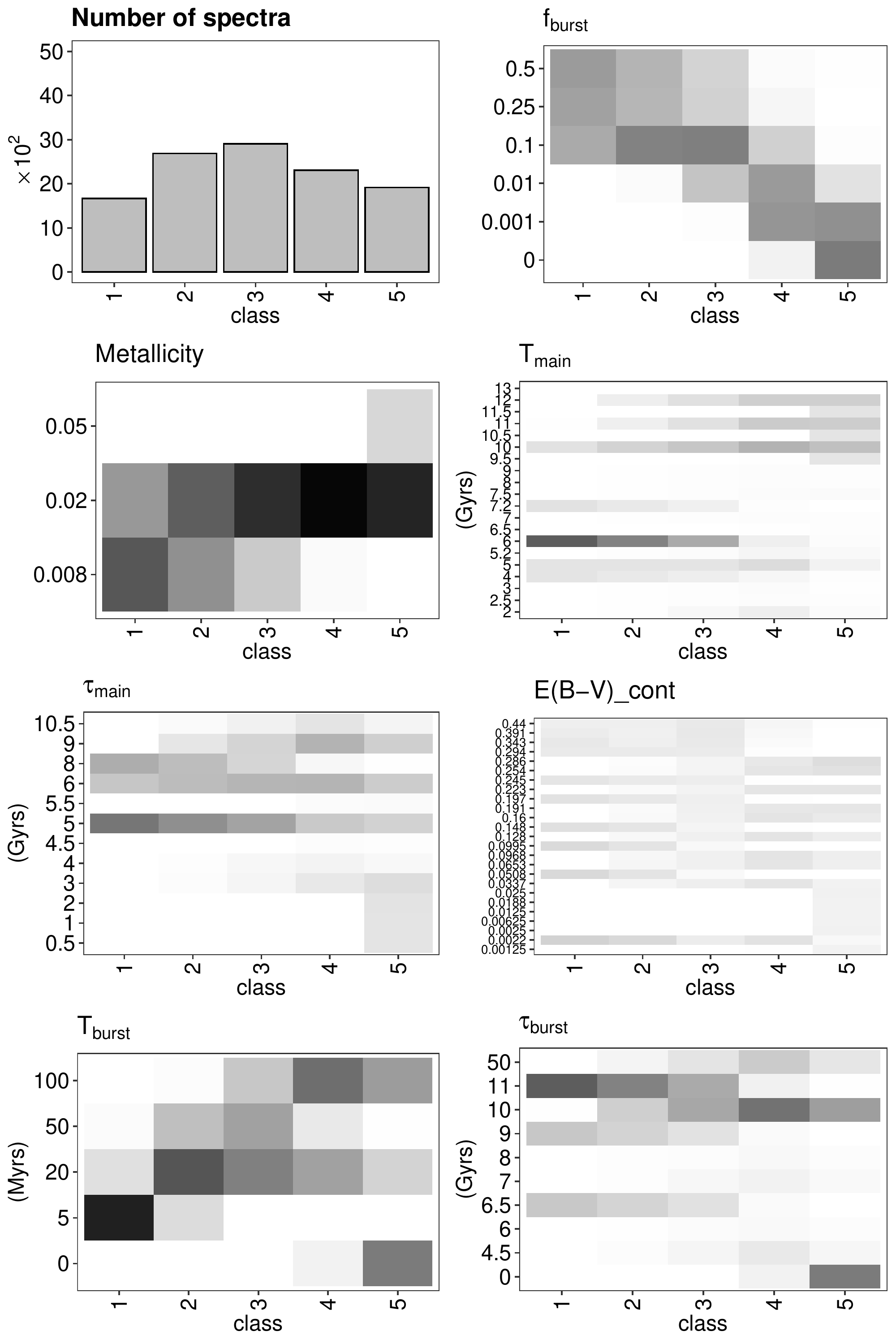}
        \caption{S/N=1 (see Fig.~\ref{fig:heatmaps_SNRINFK14}).}
\end{figure}

\begin{figure}[h!]
        \centering
        \includegraphics[scale=0.34]{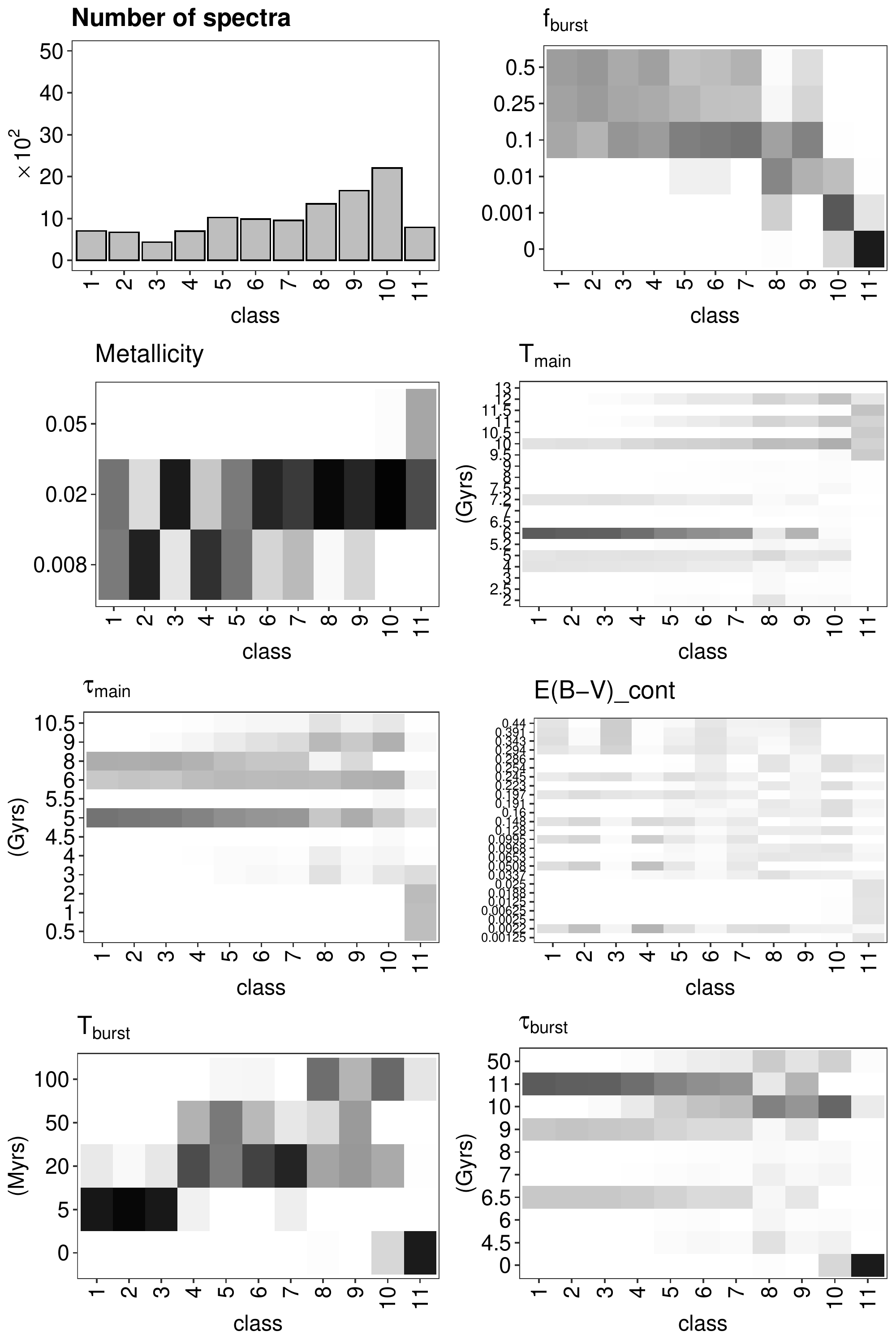}
        \caption{S/N=3 (see Fig.~\ref{fig:heatmaps_SNRINFK14}).}
\end{figure}

\begin{figure}[h!]
        \centering
        \includegraphics[scale=0.34]{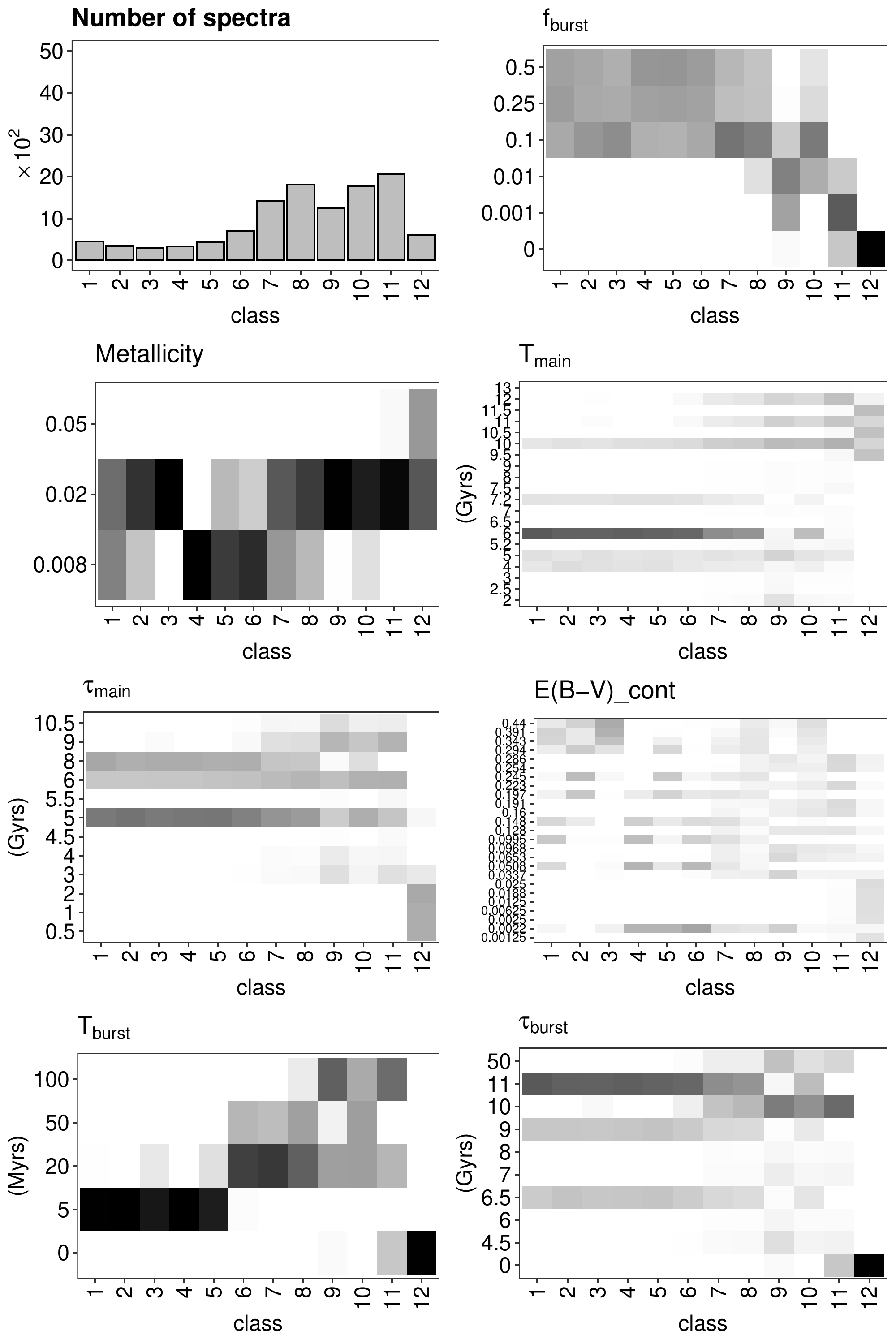}
        \caption{S/N=5 (see Fig.~\ref{fig:heatmaps_SNRINFK14}).}
\end{figure}

\begin{figure}[h!]
        \centering
        \includegraphics[scale=0.34]{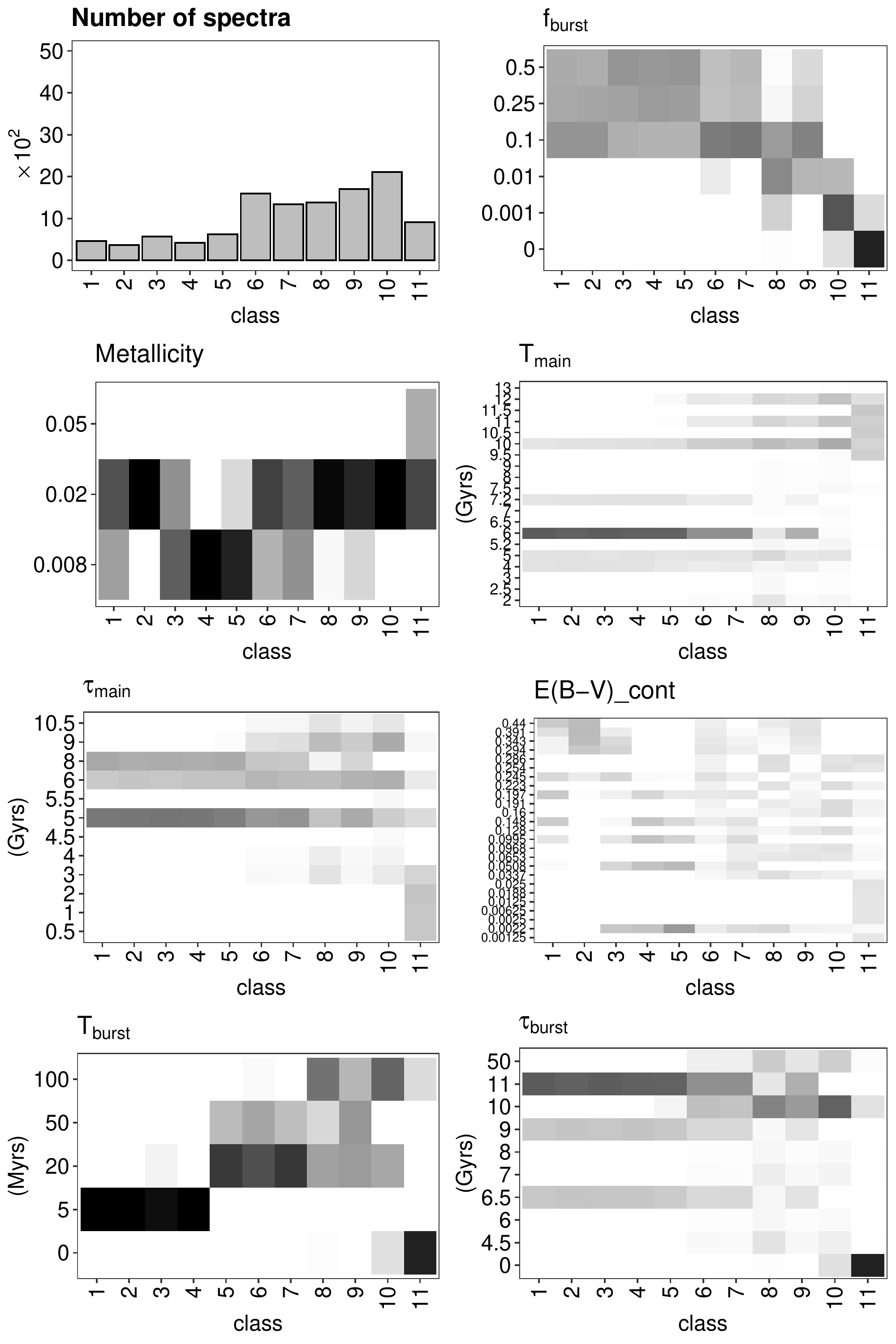}
        \caption{S/N=10 (see Fig.~\ref{fig:heatmaps_SNRINFK14}).}
\end{figure}

\begin{figure}[t!]
        \centering
        \includegraphics[scale=0.34]{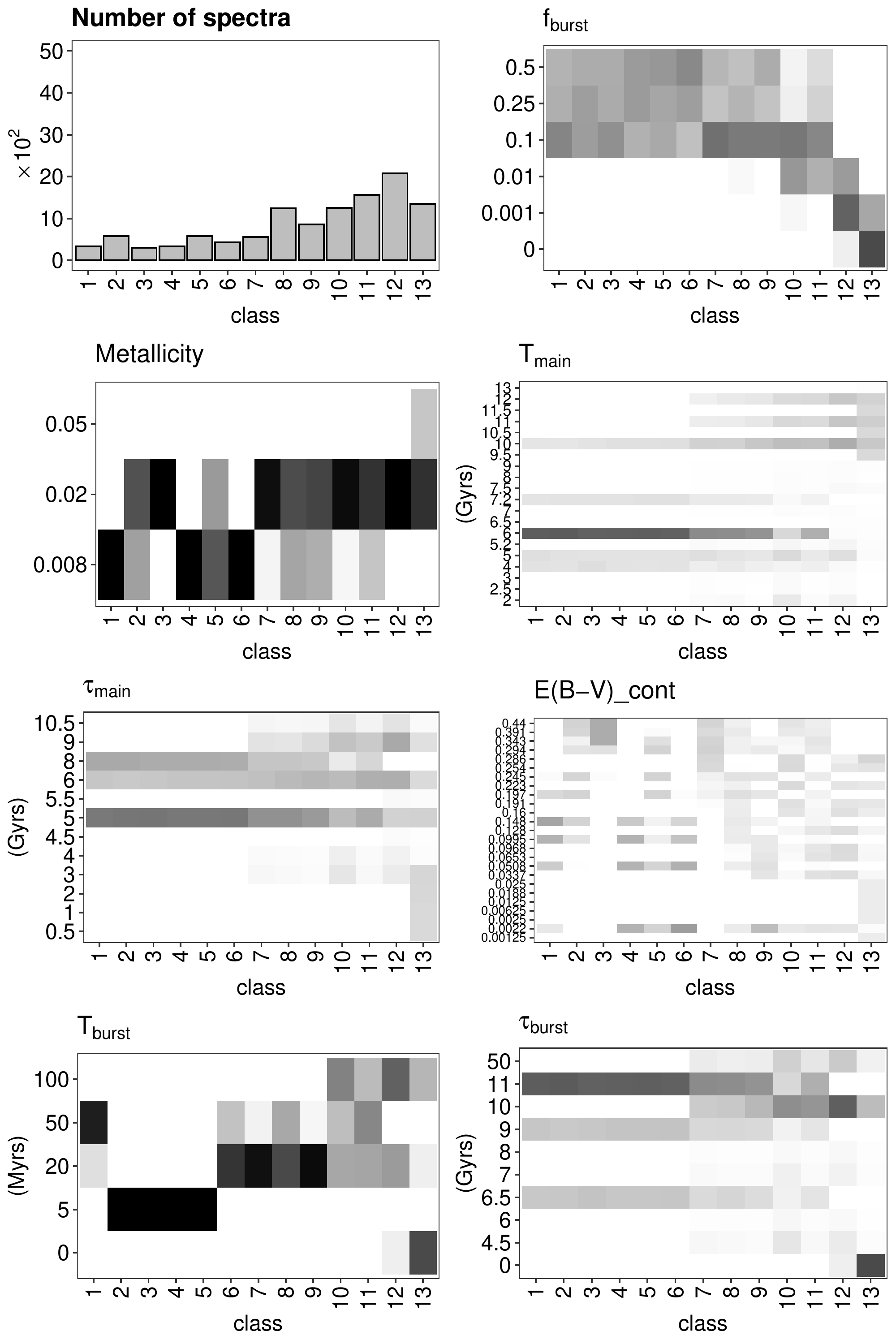}
        \caption{S/N=100 (see Fig.~\ref{fig:heatmaps_SNRINFK14}).}
\end{figure}

\subsection{LDA analysis}

\begin{figure}[H]
        \centering
        \includegraphics[scale=0.44]{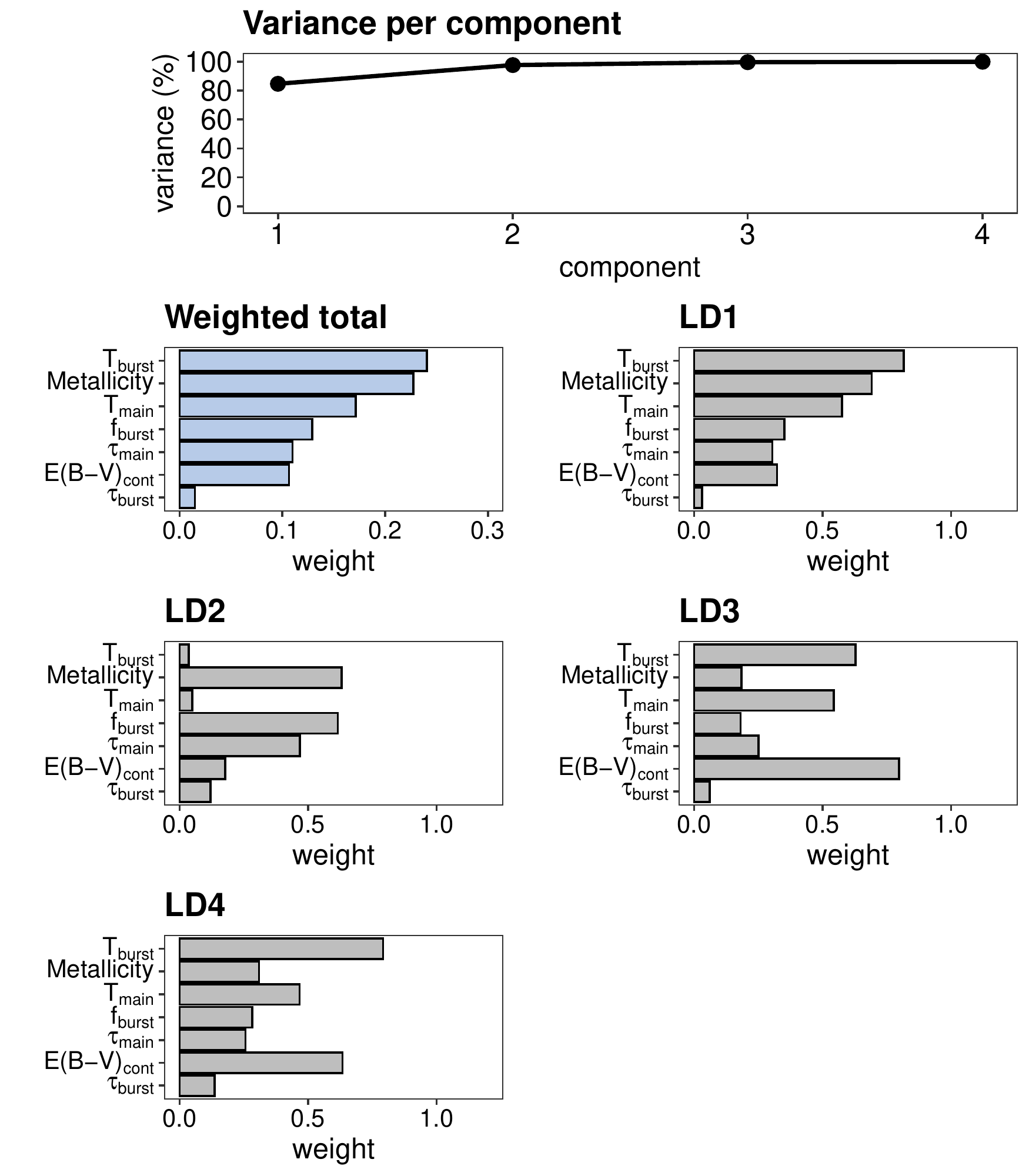}
        \caption{S/N=1 (see Fig.~\ref{fig:LDA_SNRINFK14}).}
\end{figure}

\begin{figure}[H]
        \centering
        \includegraphics[scale=0.44]{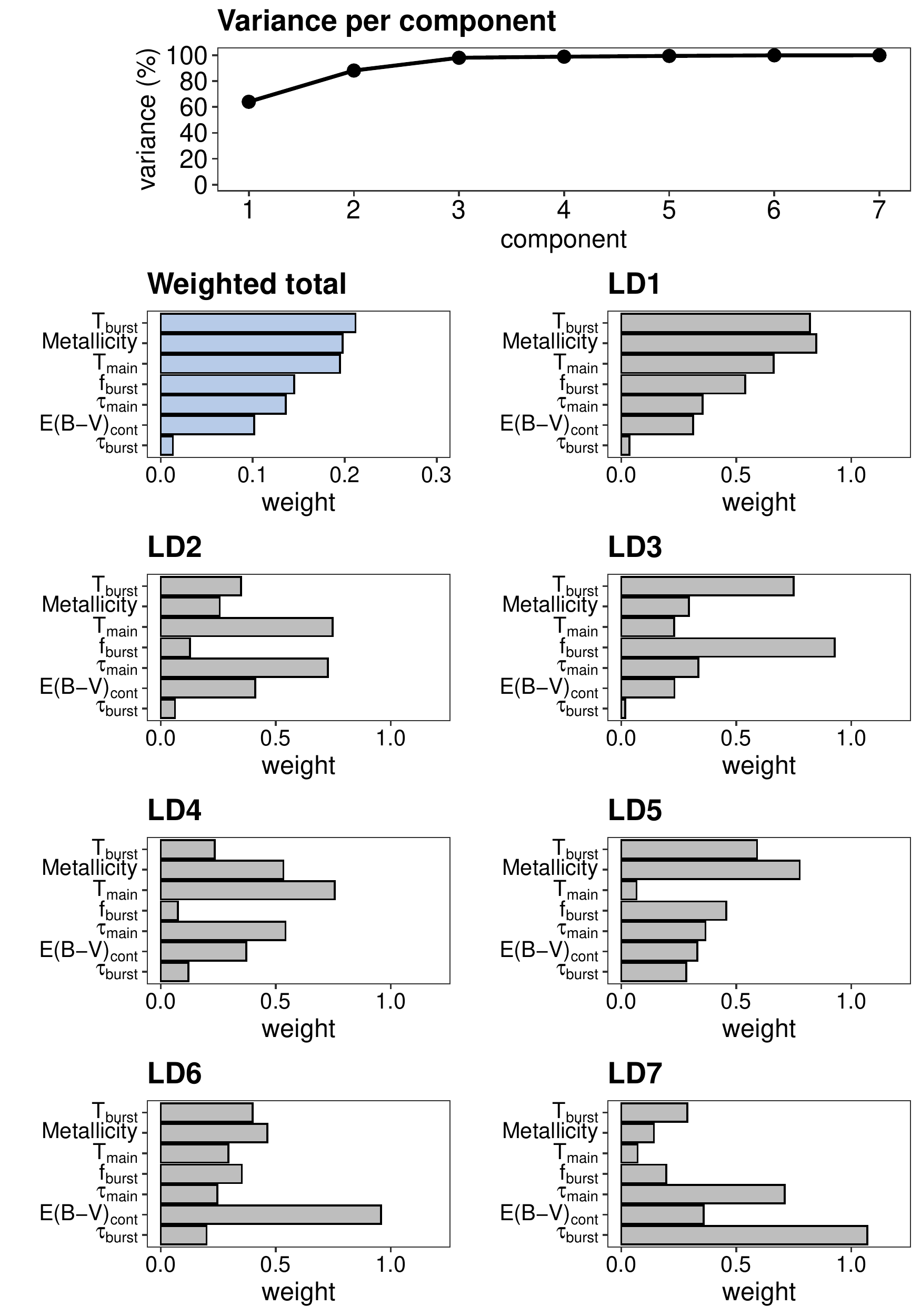}
        \caption{S/N=3 (see Fig.~\ref{fig:LDA_SNRINFK14}).}
\end{figure}

\begin{figure}[H]
        \centering
        \includegraphics[scale=0.44]{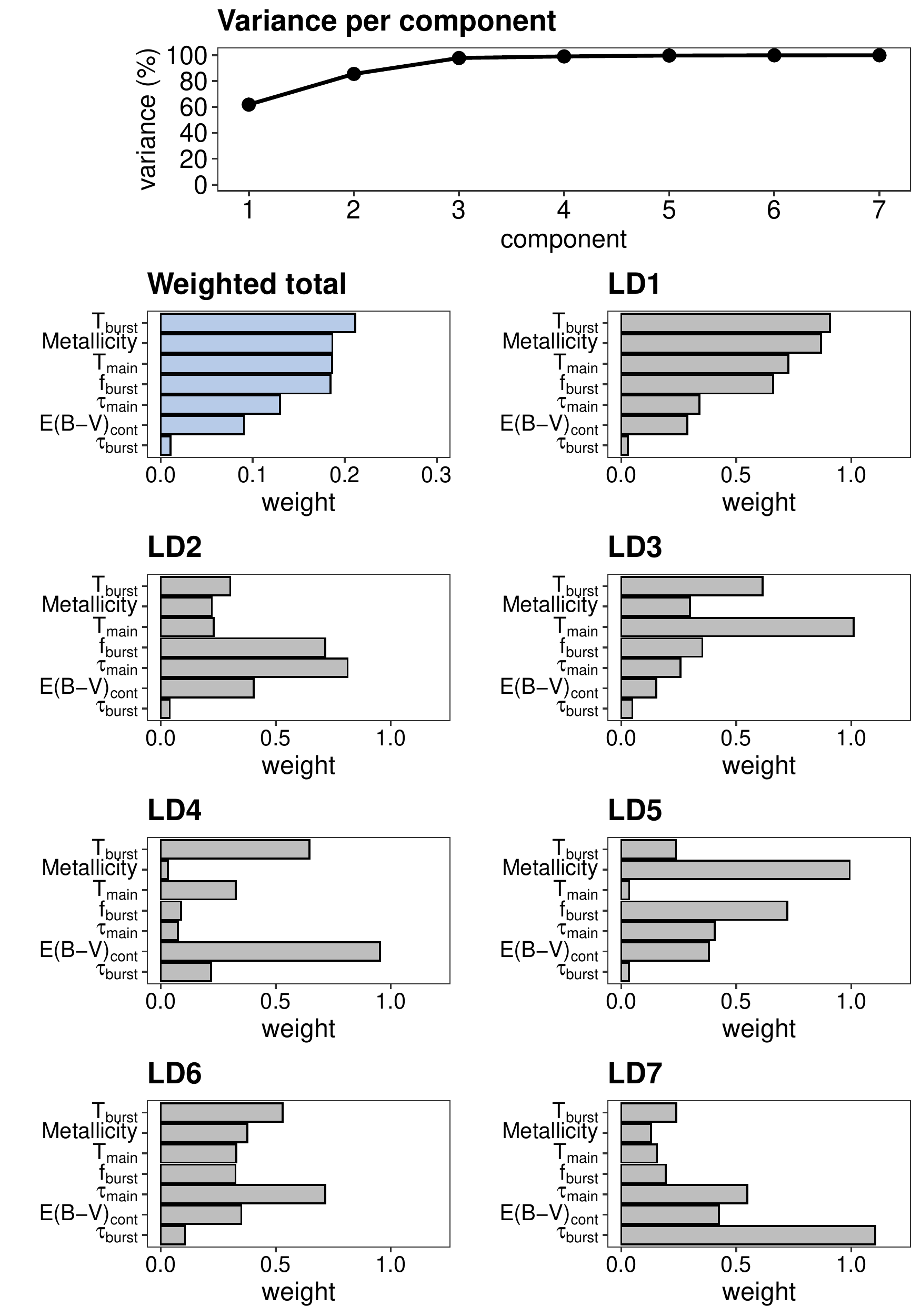}
        \caption{S/N=5 (see Fig.~\ref{fig:LDA_SNRINFK14}).}
\end{figure}

\begin{figure}[H]
        \centering
        \includegraphics[scale=0.44]{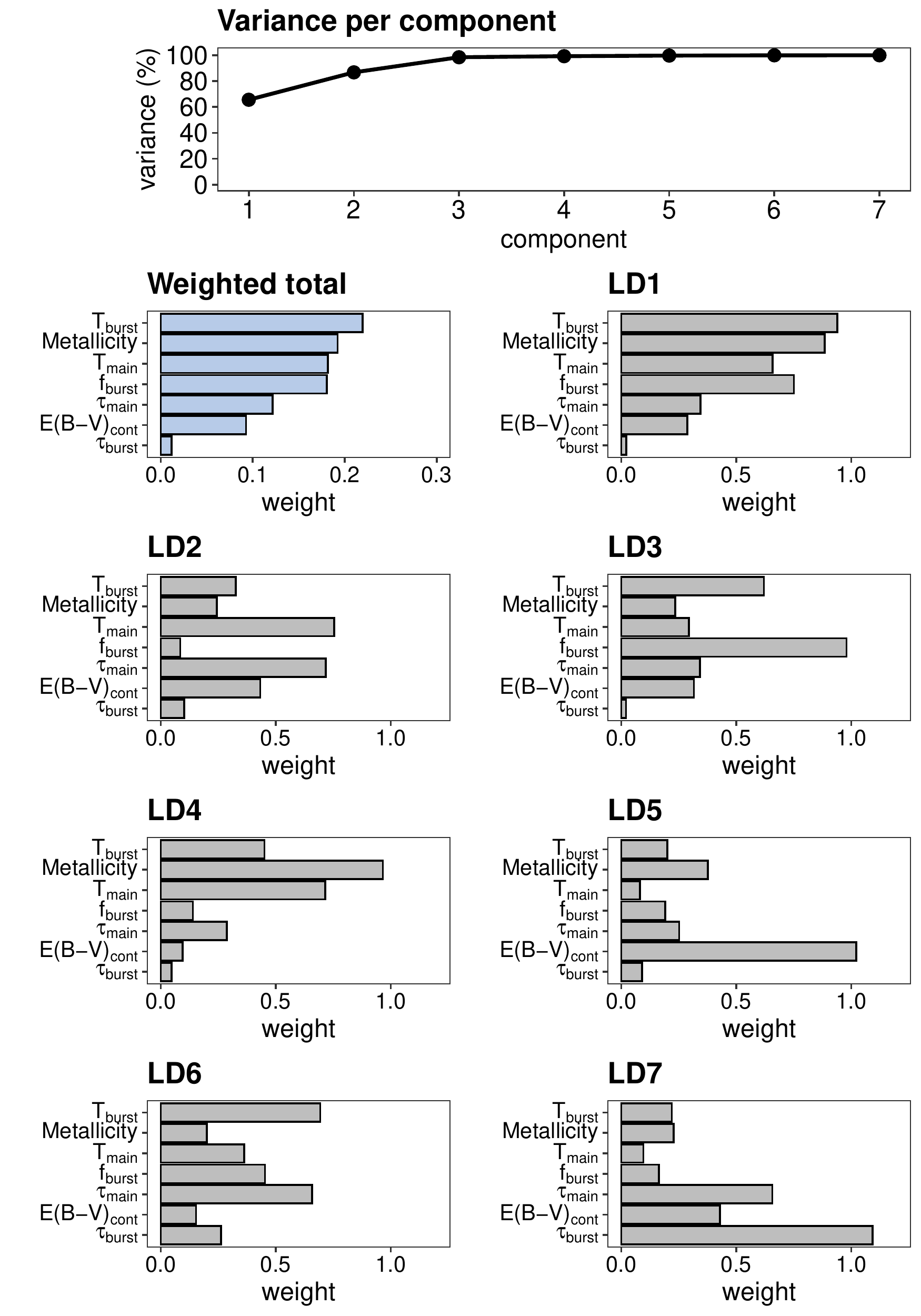}
        \caption{S/N=10 (see Fig.~\ref{fig:LDA_SNRINFK14}).}
\end{figure}

\begin{figure}[H]
        \centering
        \includegraphics[scale=0.44]{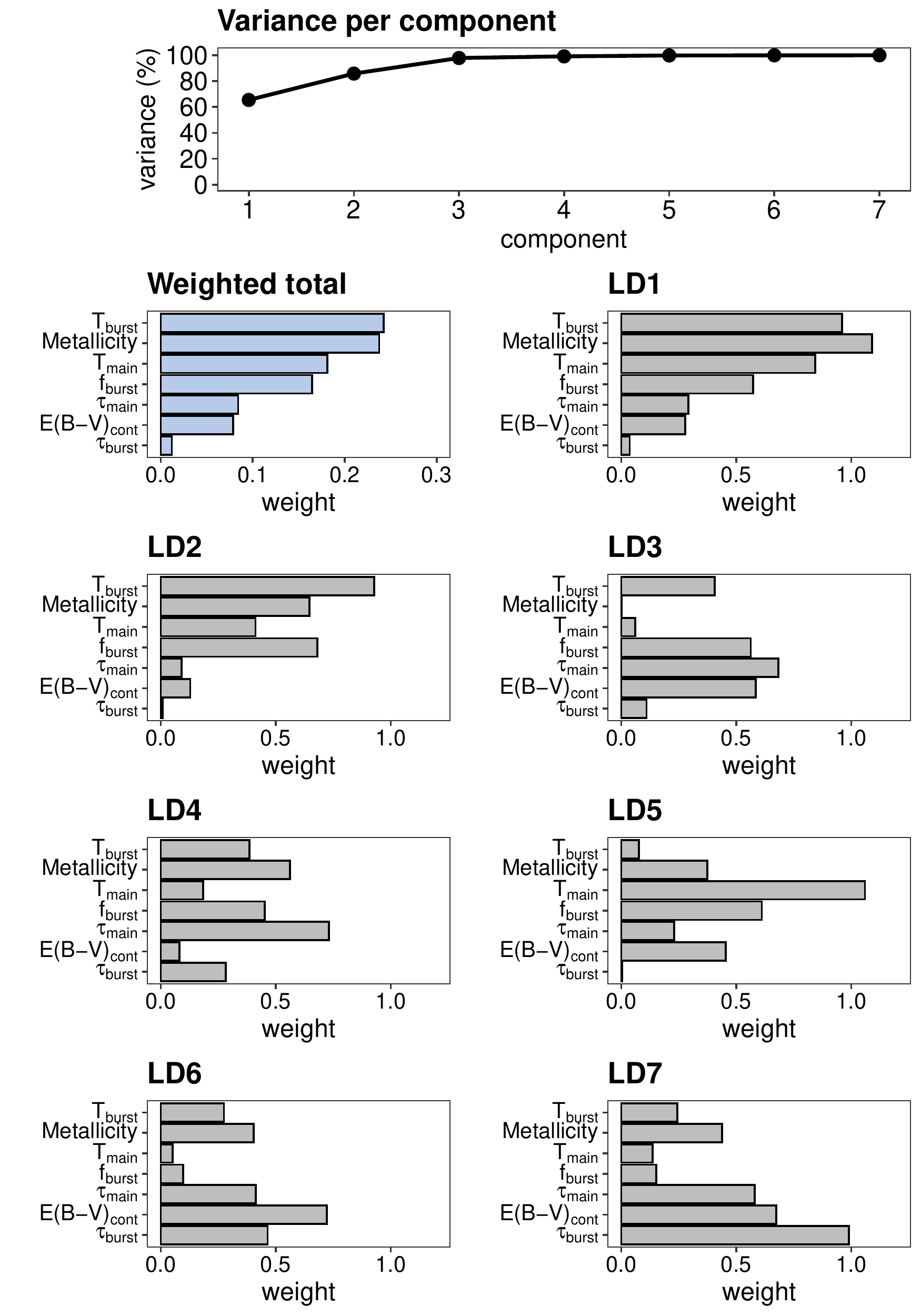}
        \caption{S/N=100 (see Fig.~\ref{fig:LDA_SNRINFK14}).}
\end{figure}

\subsection{Mean spectra}
\begin{figure}[H]
        \centering
        \includegraphics[width=\hsize]{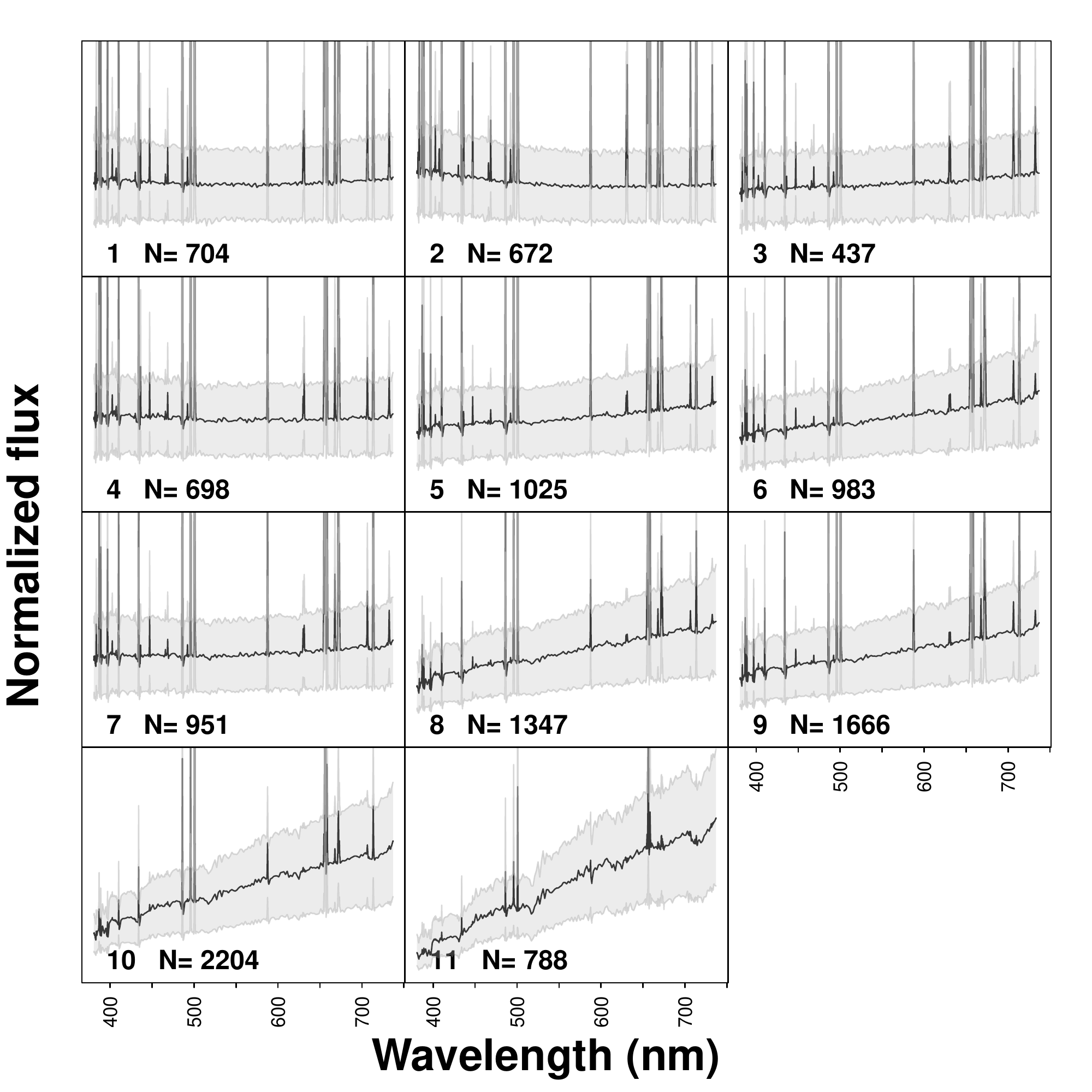}
        \caption{S/N=3 (see Fig.~\ref{fig:spectra_SNRINFK14}).}
\end{figure}

\begin{figure}[H]
        \centering
        \includegraphics[width=\hsize]{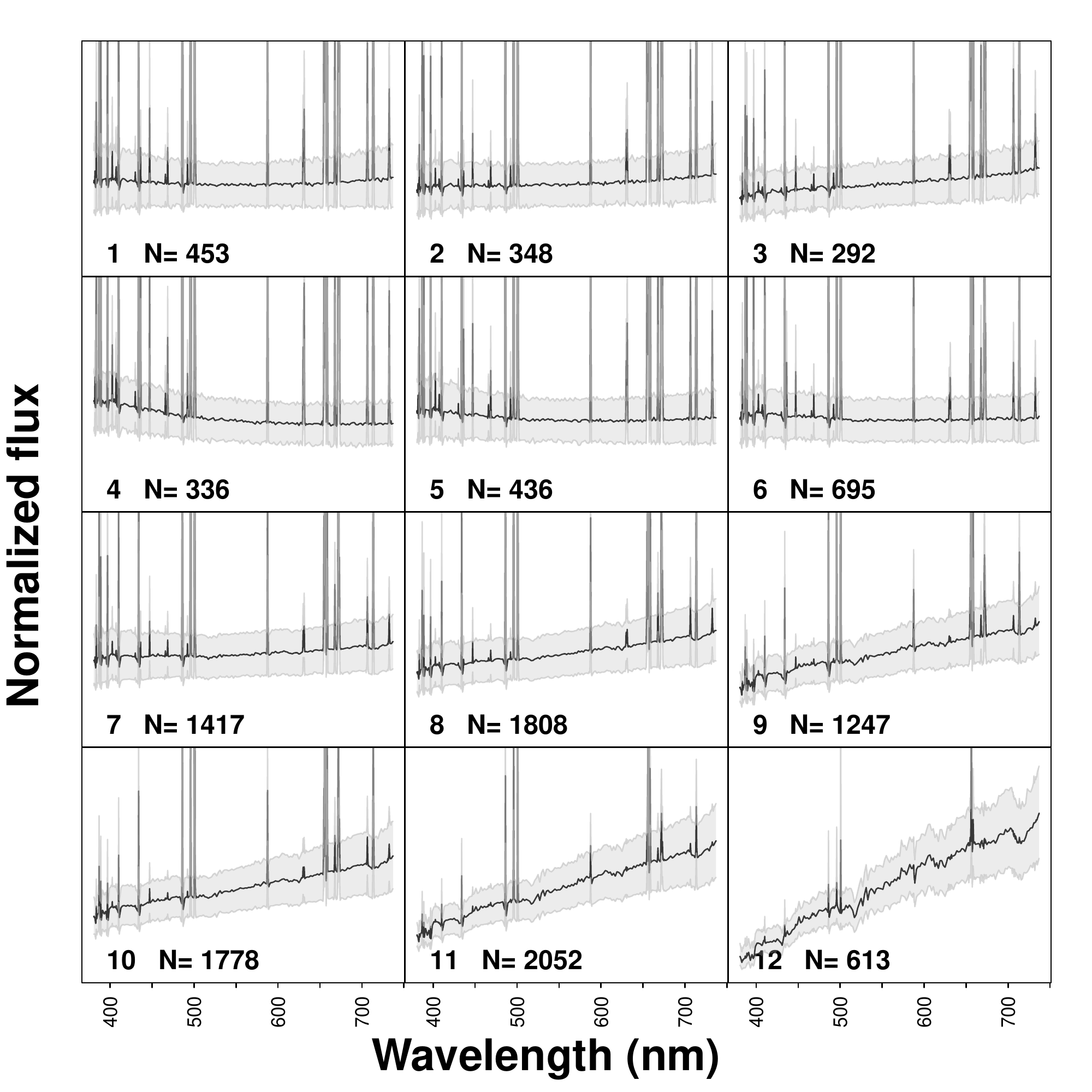}
        \caption{S/N=5 (see Fig.~\ref{fig:spectra_SNRINFK14}).}
\end{figure}

\begin{figure}[H]
        \centering
        \includegraphics[width=\hsize]{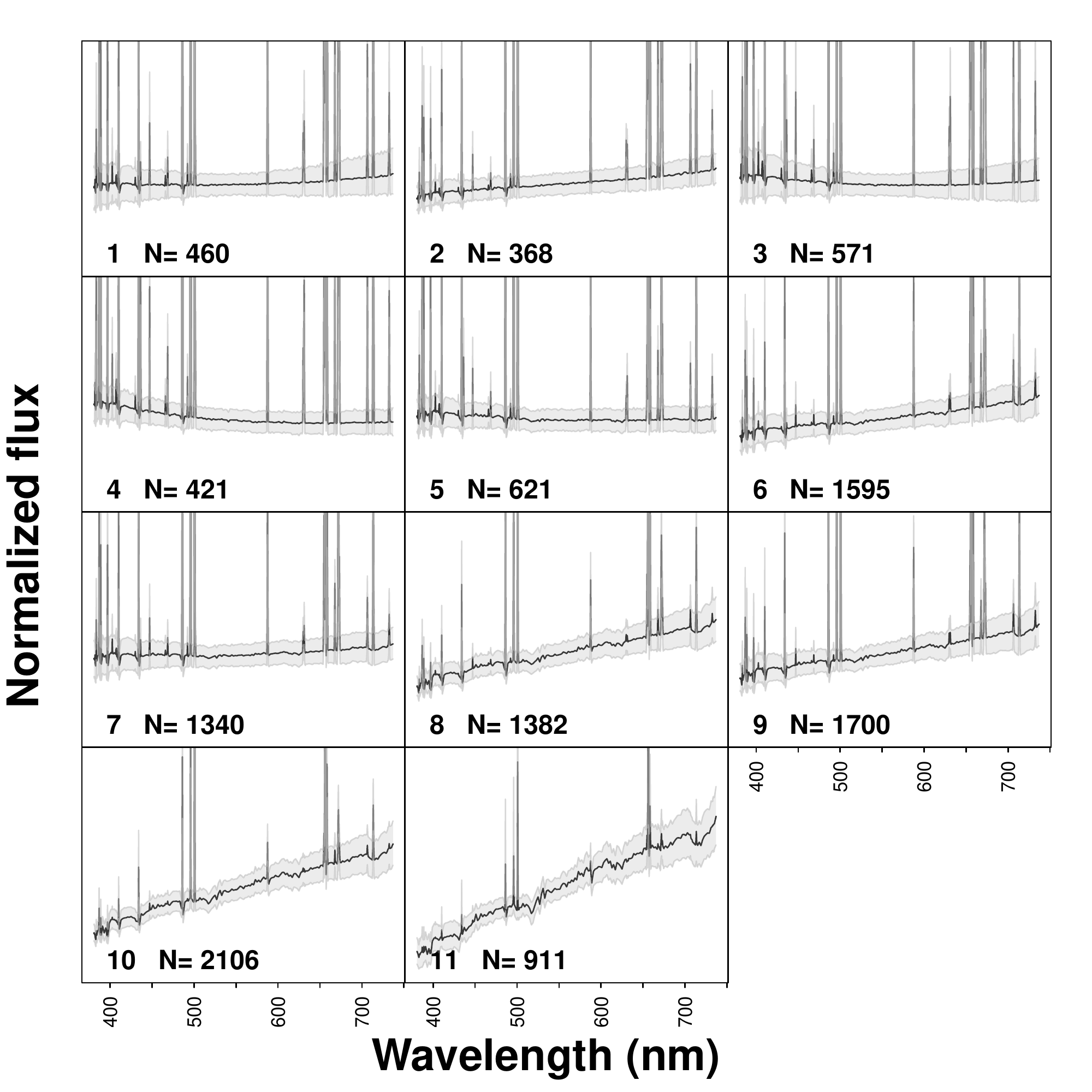}
        \caption{S/N=10 (see Fig.~\ref{fig:spectra_SNRINFK14}).}
\end{figure}

\begin{figure}[H]
        \centering
        \includegraphics[width=\hsize]{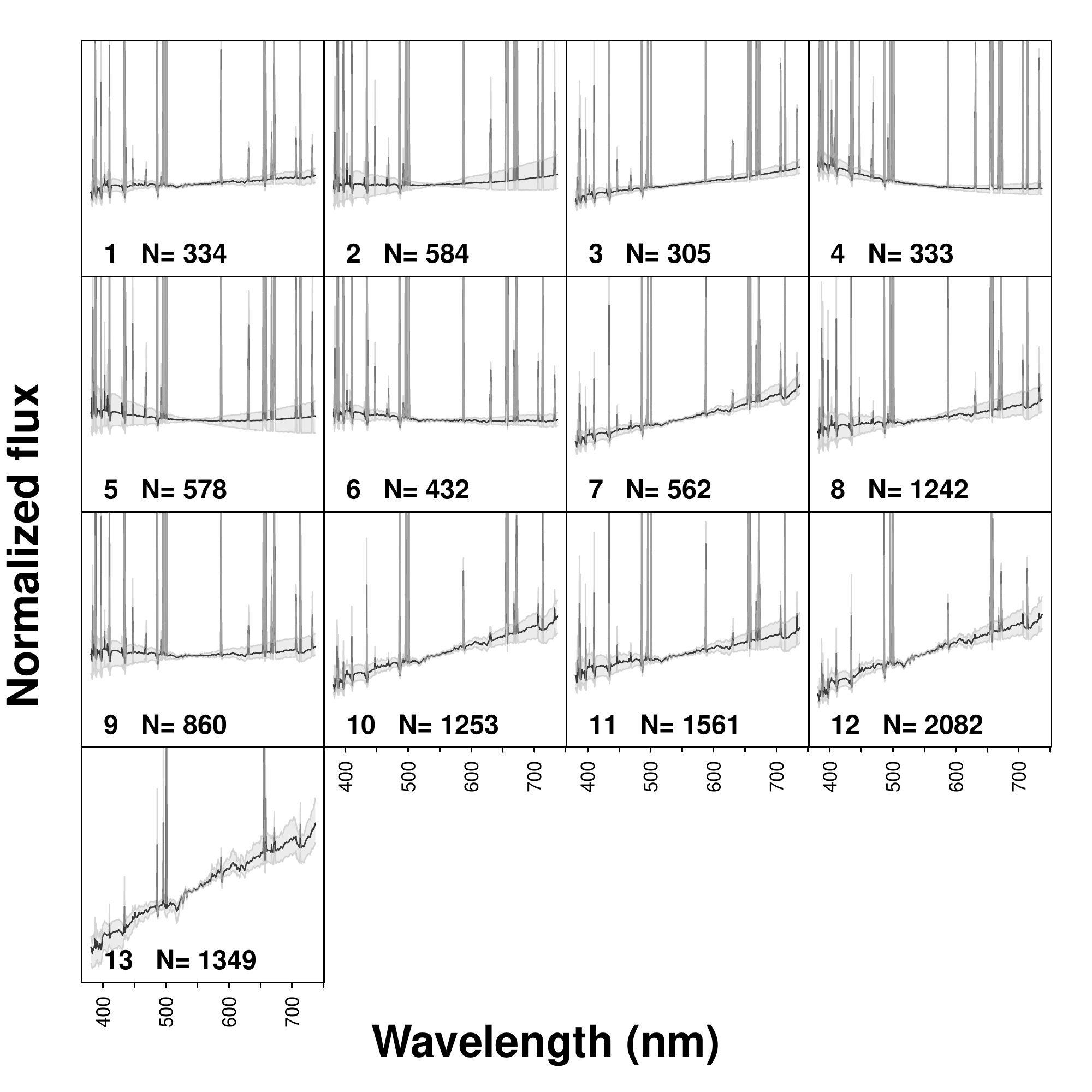}
        \caption{S/N=100 (see Fig.~\ref{fig:spectra_SNRINFK14}).}
\end{figure}

\clearpage
\section{Toy model}
\label{toymodel}

To try to visualise the conditions for the gap at K=13 in the ICL curve for the noiseless case (Sect.~\ref{section:noiseless:nbclusters} and Fig.~\ref{ICL_SNRINF}), we constructed a toy model by trial and error. The simplest sample that can be built to reproduce this behaviour is a sample of five variables and 1000 observations to ensure a perfect reproducibility. 

We consider the following matrix made with the following five variables:

$$\begin{array}{ll}
\mathrm{Var1}[1:500] = 1 & \mathrm{Var1}[501:1000] = 2 \\
\mathrm{Var2}[1:300] = \mathcal{N}(10,0.01) & \mathrm{Var2}[301:1000] = \mathcal{N}(15,0.01) \\
\mathrm{Var3}[1:800] = 1 & \mathrm{Var3}[801:1000] = 2 \\
\mathrm{Var4}[1:500] = \mathcal{N}(100,0.01) & \mathrm{Var4}[501:1000] = \mathcal{N}(150,0.05) \\
\mathrm{Var5}[1:200] = 1 & \mathrm{Var5}[201:1000] = \mathcal{N}(4,0.1)
\end{array}
$$

where VarX$[i:j]$ designates the indices from $i$ to $j$ of variable VarX, and $\mathcal{N}(\mu,\sigma^2)$ means that the values are drawn from a normal distribution  of mean $\mu$ and standard deviation $\sigma$.

\begin{figure}[b]
        \centering
        \includegraphics[width=0.9\linewidth]{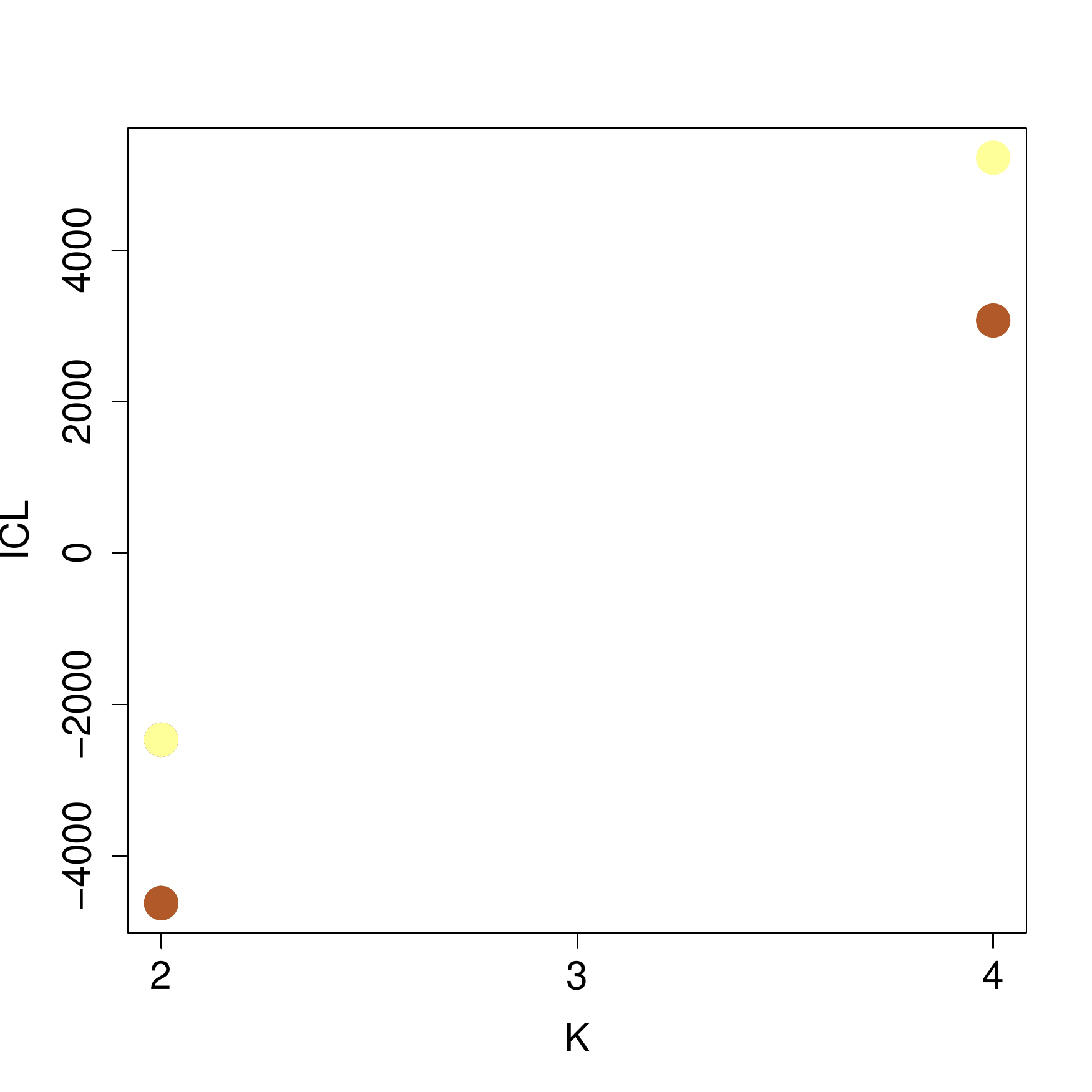}
        \caption{All the ICL values as a function of the number of cluster K obtained for the toy model. Each point corresponds to a successful run of \textit{Fisher-EM} for one of the 12 statistical models. This figure should be compared with Fig.~\ref{ICL_SNRINF}.}
        \label{fig:toymodelICL1}
\end{figure}

\begin{figure}
        \centering
        \includegraphics[width=0.9\linewidth]{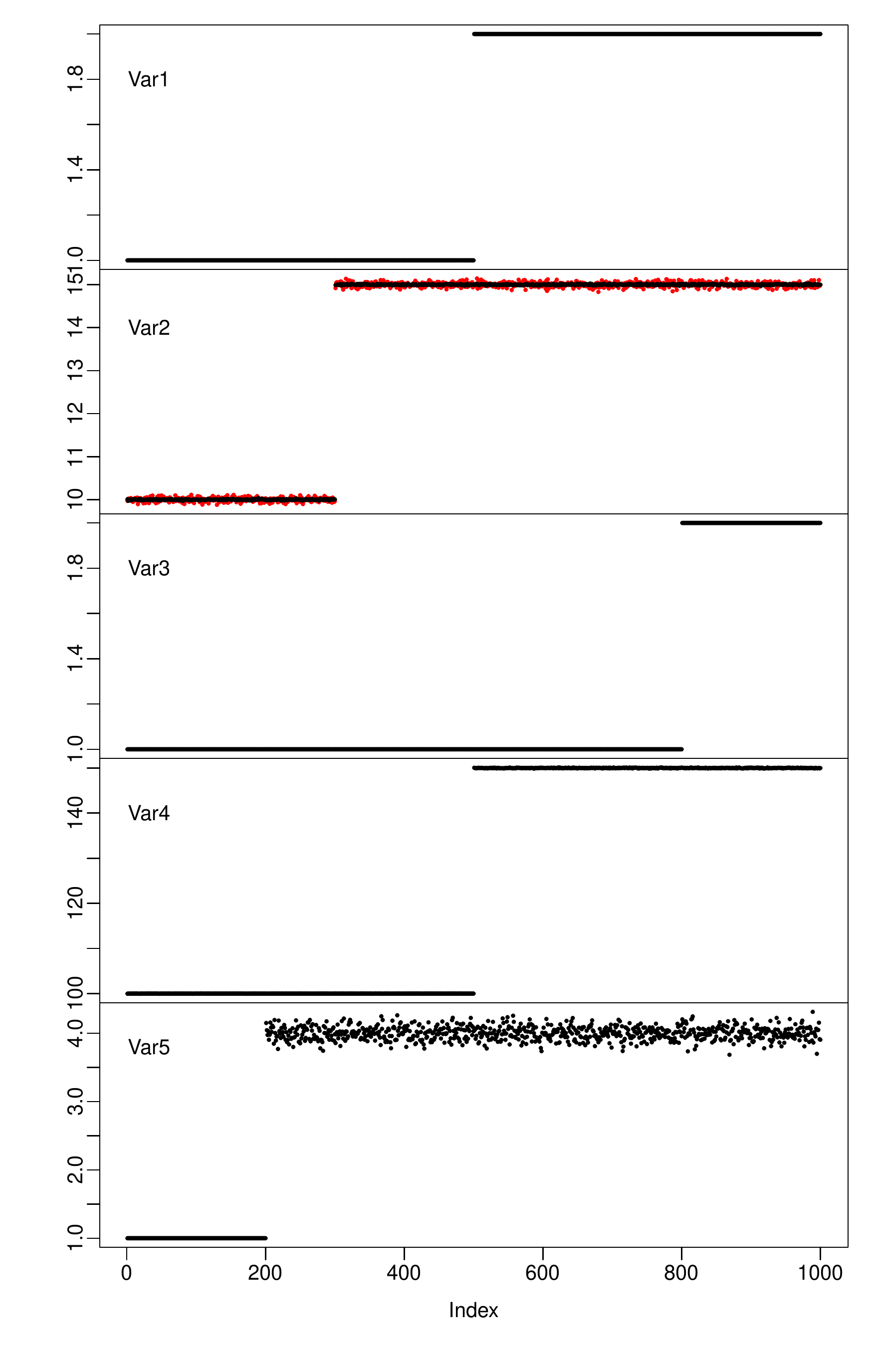}
        \caption{Values of the five variables for the toy model with 1000 observations that yield the ICL curve in Fig~\ref{fig:toymodelICL1}. The points in red in the second panel (Var2) show a slightly increased dispersion that yields the ICL curve in Fig~\ref{fig:toymodelICL2}.}
        \label{fig:toymodel}
\end{figure}

This sample (Fig~\ref{fig:toymodel}) yields an ICL curve Fig~\ref{fig:toymodelICL1} with a gap at K=3 (\textit{Fisher-EM} never converges) and a much higher value at K=4 than at K=2. This behaviour is identical to the one at K=13 in Fig.~\ref{ICL_SNRINF}.

\begin{figure}
        \centering
        \includegraphics[width=0.9\linewidth]{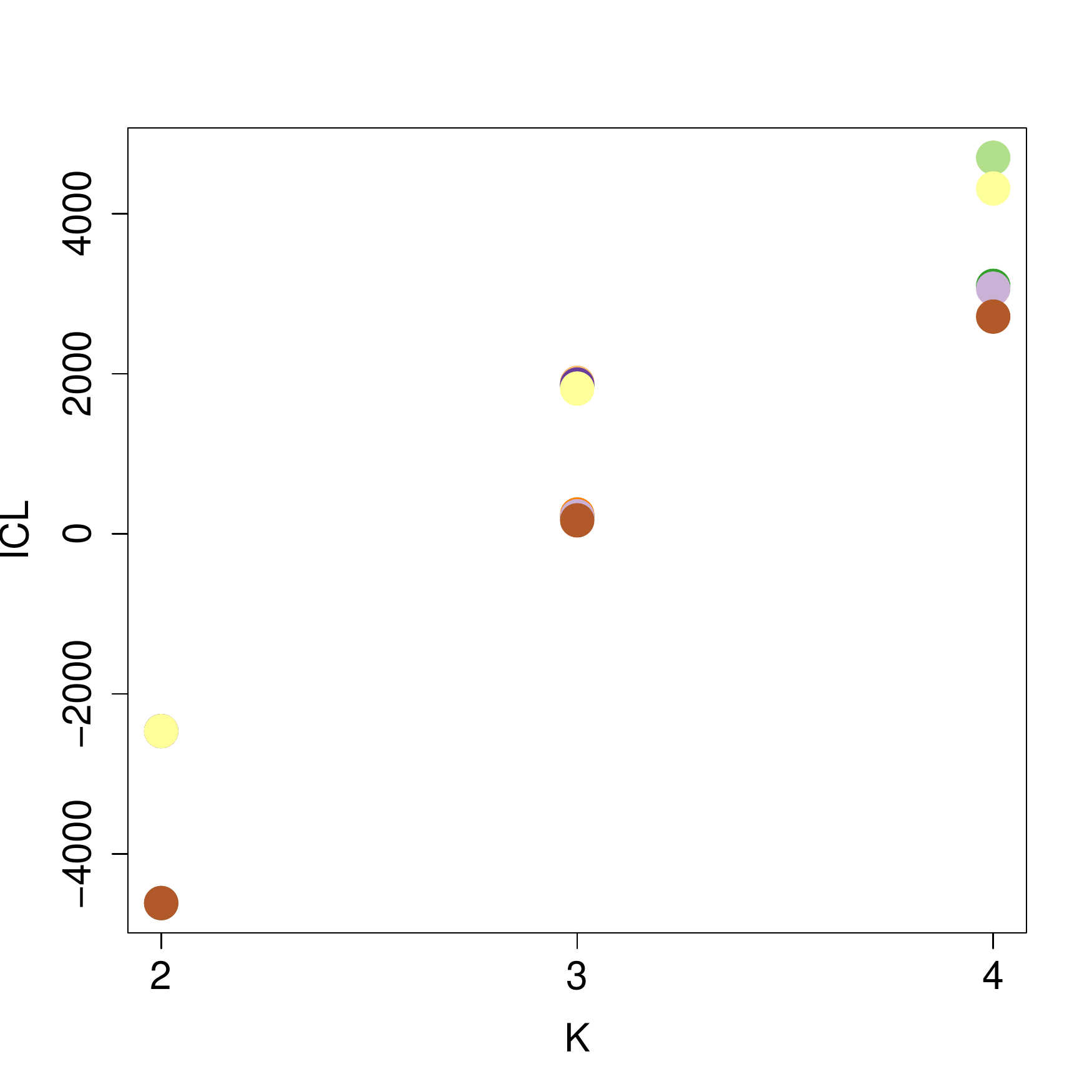}
        \caption{Same as Fig.~\ref{fig:toymodelICL1} with the slightly mode dispersed variable Var2.}
        \label{fig:toymodelICL2}
\end{figure}

Adding some dispersion in Var2 by increasing $\sigma^2$ from 0.01 to 0.05,
$$\begin{array}{ll}
 \mathrm{Var2}[1:300] = \mathcal{N}(10,0.05) & \mathrm{Var2}[301:1000] = \mathcal{N}(15,0.05) 
\end{array}$$
as represented by the red points in Fig~\ref{fig:toymodel}, the \textit{Fisher-EM} analysis always yields a solution (Fig~\ref{fig:toymodelICL2}).

This behaviour is thus very similar to the one obtained on the CIGALE sample, and is thus explained by the very peculiar distribution of the observations in the multivariate data space.

\end{appendix}

\end{document}